\documentclass[twocolumn,showpacs,preprintnumbers, 
               superscriptaddress,eqsecnum]{revtex4}
\usepackage{graphicx, fancybox}
\usepackage{amsmath,amssymb}

\newcommand{\bm}[1]{\mbox{\boldmath$#1$}}
\newcommand{\vect}[1]{\mathbf{#1}}
\newcommand{\abs}[1]{\left|#1\right|}

\newcommand{\im}{\mathrm{Im}}
\newcommand{\re}{\mathrm{Re}}

\newcommand{\Slash}[1]{\ooalign{\hfil/\hfil\crcr$#1$}}

\newcommand{\ef}{E_f}
\newcommand{\eb}{E_b}

\newcommand{\vk}{\vect{k}}
\newcommand{\Tr}{\mathrm{Tr}}

\newcommand{\po}[2]{\Lambda_#1(#2)}

\newcommand{\mb}{m_b}
\newcommand{\mf}{m_f}
\newcommand{\dm}{M_{-}}
\newcommand{\tm}{M_{+}}

\newcommand{\imsigpvac}{\im \Sigma^{R,T=0}_+}

\begin{document}

\preprint{YITP-07-74}

\title{
Spectral properties of massless and massive quarks
coupled with massive boson \\
at finite temperature
}

\author{Masakiyo Kitazawa}
\email{kitazawa@phys.sci.osaka-u.ac.jp}
\affiliation{
Department of Physics, Osaka University, Toyonaka, Osaka 560-0043, Japan}

\author{Teiji Kunihiro}
\email{kunihiro@yukawa.kyoto-u.ac.jp} 
\affiliation{
Yukawa Institute for Theoretical Physics, Kyoto University, Kyoto,
606-8502, Japan}

\author{Kazuya Mitsutani}
\email{kazuya@yukawa.kyoto-u.ac.jp}
\affiliation{
Yukawa Institute for Theoretical Physics, Kyoto University, Kyoto,
606-8502, Japan}

\author{Yukio Nemoto}
\email{nemoto@hken.phys.nagoya-u.ac.jp}
\affiliation{
Department of Physics, Nagoya University, Nagoya, 464-8602, Japan}

\begin{abstract}
We study the  properties of  massless and massive
quarks coupled with a scalar and pseudoscalar boson 
at finite temperature in Yukawa models at the one-loop order.
The behavior of the spectral function and the pole structure
of the propagator are analyzed as functions of temperature $T$
and the quark mass $m_f$.
It is shown that the three-peak structure of the spectral function
found in a previous work for massless quarks 
is formed at 
temperatures comparable to the boson mass even for finite $m_f$,
but gradually ceases to exist as $m_f$ becomes larger.
We identify the   three poles of the quark propagator 
corresponding to the collective excitations of the quark
in the complex energy plane.
It is shown that 
the three trajectories made by  the poles along with 
a variation of $T$ undergo 
a structural rearrangement at a 
critical quark mass when $m_f$ is increased.
This suggests that the physics content of the collective quark
excitations is changed in a drastic way at this point.
The results are nicely accounted for with 
the notion of the level mixing 
induced by a resonant scattering of the massive boson
with quarks and holes of thermally excited anti-quarks. 

\end{abstract}

\date{\today}

\pacs{11.10.Wx, 11.30.Rd, 12.38.Bx, 12.38.Mh, 25.75.Nq}
\maketitle

\section{Introduction}
\label{sec:intro}

Hadronic matter undergoes deconfinement and chiral phase
transitions to the quark-gluon plasma (QGP) phase
at finite temperature ($T$).
Owing to the asymptotic freedom of QCD,
the QGP phase at asymptotically high $T$
is composed of approximately free quarks and gluons,
where a perturbation expansion is valid.
It is known that the leading order of such an expansion is obtained
with the hard thermal loop (HTL) approximation \cite{HTL} which
enables us to calculate the quark and gluon propagators
in a gauge-invariant way.
One of notable features obtained in the HTL approximation is that
the quarks and gluons have collective excitations 
with mass gaps called thermal masses proportional to $gT$, 
where $g$ is the gauge coupling \cite{plasmino,LeBellac}.

Recently, properties of the QGP phase 
near the critical temperature ($T_c$) acquire much interest.
Measurements of the elliptic flow of hadrons in the
heavy-ion collisions were made at the Relativistic Heavy Ion Collider (RHIC),
and the subsequent phenomenological analyses 
have shown that the experimental results 
are in remarkably good agreement with the predictions of 
 the ideal fluid dynamics \cite{RHIC}.
This result 
suggests that the created matter is a strongly coupled system.
The present paper is concerned with the properties of  quarks
in such a system, 
which constitute the basic degrees of freedom of the created 
matter in the deconfined phase together with gluons.
We explore possible quasi-particle picture 
of quarks  in the QGP phase near $T_c$.

It is not a priori clear  whether 
quarks manifest themselves as 
well-defined quasi-particles in such a nonperturbative region;
quasi-particles in Landau's sense correspond to peaks with a small  
width in the
spectral function with relevant quantum numbers.
In the phenomenological side,
 the success of the recombination model to describe
the RHIC data \cite{RECO} 
tells us that the quark quasi-particles play an important role in 
some stage of the evolution of the created matter 
in RHIC.
In the theoretical side,
a recent lattice QCD simulation \cite{Karsch:2007wc}  
indicates the existence of the quasi-particles of quarks 
with small decay width, though in quenched approximation.
Quasi-particle properties of quarks
are also obtained in a strong
coupling gauge theory based on the Schwinger-Dyson 
equation at finite $T$, unless the coupling is too large
\cite{Harada:2007gg}.
The quarks thus will exist as one of the basic
degrees of freedom of the system even near $T_c$,
and hence revealing their nature will provide us 
with important information necessary to understand 
the nature of QGP phase at the strong coupling.

To study the quasi-particle properties of quarks
around $T_c$, 
we notice the possible existence of hadronic excitations  even in the
QGP phase near $T_c$ \cite{J/Psi,HK85}.
The numerical results in lattice QCD suggest the existence of 
such states in the heavy-quark sector \cite{Lattice}.
The existence of the mesonic excitations in the light-quark sector 
was also predicted as being the soft modes associated with
the chiral transition \cite{HK85}.
If such bosonic modes are light enough so that
many particles are thermally excited in the system, 
they can in turn strongly affect excitation spectra of quarks.
Part of the authors in the present paper (M.K., T.K. and Y.N.) considered such 
possibility in Refs.~\cite{KKN1,KKN2}.
They first evaluated the effect of the chiral soft modes
on quarks in a chiral effective model \cite{KKN1}
where the soft modes are described as composite bosons
 composed of a quark and anti-quark: An interesting result 
obtained there was that 
there appears a novel three-peak structure 
in the quark spectral function near but above $T_c$.
It was also pointed out that the coupling of
quarks with bosonic excitations which have a finite mass and
small width 
is essential for the formation of the multi-peak structure.
In fact, 
it was  later shown \cite{KKN2} that
the three-peak structure in the quark spectral function 
emerges at intermediate temperatures even 
in the Yukawa models composed of a massless fermion
and a massive boson with a zero width, 
irrespective whether the boson is of scalar, pseudoscalar, vector or
axial-vector type.

In the previous work \cite{KKN1,KKN2}, the massless quarks
were exclusively adopted
to explore the effects of the soft modes on the quarks
in the most ideal case where the chiral symmetry is exact.
In reality,
the quarks in the QGP phase near $T_c$
should have current or dynamical  Dirac masses
which may be as large as the order of $T_c$ and 
may not be negligible.
In this paper, we examine how
the finite Dirac mass of the quark modifies the spectral
properties, employing the Yukawa model composed of
a massive quark with a small mass and a massive boson.
We shall show that 
the three-peak structure found in Ref.~\cite{KKN2}
survives even with finite $m_f$, although
one of the three peaks is gradually 
suppressed as $m_f$ is larger.

In addition to the analysis of the spectral function,
we newly explore the pole structure of the quark propagator
in the complex-energy plane and their residues.
We show that there always exist three poles corresponding
to the peaks in the spectral function
in the complex-energy plane at finite $T$.
It is found that the trajectories of these poles 
along  with an increase of $T$ 
undergoes a structural change at a critical value of $m_f$
as $m_f$ is increased.
We give a physical interpretation for these
properties of the quark spectrum at finite $m_f$
adapting the notion of 
the level repulsion in quantum mechanics.

This paper is organized as follows.
In Secs.~\ref{sec:model}-\ref{sec:polesc},
we study the spectral properties of the quark
coupled with a scalar boson.
After introducing the Yukawa model in Sec.~\ref{sec:model}, 
we present the numerical result for the quark spectral function
in Sec.~\ref{sec:spcsc}
and the pole structure of the propagator in Sec.~\ref{sec:polesc}.
We then discuss the spectral properties of the quark
coupled with a pseudoscalar boson in Sec.~\ref{sec:PS}.
In Sec.~\ref{sec:level-mixing},
we elucidate the physical mechanism
of the $m_f$-dependence of the quark spectrum.
Section~\ref{sec:summ} is devoted to the summary and 
some discussions.
In Appendix \ref{app:renormalization},
we present the renormalization procedure of the quark propagator 
using
the subtracted dispersion relations.
In Appendix \ref{app:3peak}, we introduce an analytic toy model
and elucidate 
that a two-peak structure of the quark self-energy
is essential for the appearance of the
three-peak structure in the quark spectrum.
To simplify the analysis,
we restrict the following calculations to zero momentum 
to concentrate on the $T$ dependence.


\section{Yukawa model coupled with a scalar boson}
\label{sec:model}

In this section, we formulate a Yukawa model with
a massive quark and a massive scalar boson and evaluate
the quark self-energy at one loop.
The quark spectral function and some other quantities
needed in the following sections are also introduced.

We start from the following Lagrangian composed of
a quark and a scalar boson,
\begin{equation}
\mathcal{L}=\bar{\psi}(i\Slash{\partial}-\mf)\psi
+\frac{1}{2}\left[(\partial_\mu \phi)^2-\mb^2\right]
-g\bar{\psi}\phi\psi,
\label{eq:lag}
\end{equation}
with the quark mass $m_f$, the boson mass $m_b$ and
the coupling constant $g$.
If the boson is a pseudoscalar, $g$ should be replaced
with $i\gamma_5 g$.
This replacement may lead to non-trivial differences
because the quarks are massive.
The pseudoscalar case will be  analyzed in Sec.~\ref{sec:PS}.

\begin{figure}[tn]
\includegraphics[clip,width=4cm]{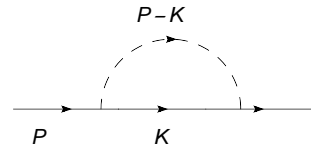}
\caption{The one-loop quark self energy.
The solid and dashed lines represent the quark
and the scalar boson, respectively.
4-momenta $P$ and $K$ denote
$P=(\bm{p}=\bm{0},\omega_m)$ and $K=(\bm{k},\omega_n)$,
respectively.
}
\label{fig:1-loop}
\end{figure}

The spectral properties of the quark 
at zero momentum can be
extracted from the retarded quark propagator 
\begin{align}
G^{\rm R}(\omega)
&=\left[(\omega+i\eta)-\mf-\Sigma^{\rm R}(\omega)\right]^{-1},
\label{eq:unprojectedgreefunction}
\end{align}
with the retarded self-energy $\Sigma^{\rm R}(\omega)$.
We calculate the quark self-energy at finite $T$ 
at one loop, which is diagrammatically
shown in Fig.~\ref{fig:1-loop}.
In the imaginary-time formalism, it is given by
\begin{equation}
\tilde{\Sigma}(i\omega_m)
=-g^2T\sum_n\int\frac{d^3k}{(2\pi)^3}
\mathcal{G}_0(\vk,i\omega_n)\mathcal{D}(-\vk,i\omega_m-i\omega_n),
\label{eq:sigma}
\end{equation}
where $\mathcal{G}_0(\vk,i\omega_n)
=[i\omega_n\gamma^0-\vk\cdot\vec{\gamma}-\mf]^{-1}$ and
$\mathcal{D}(\vk,i\nu_n)=[(i\nu_n)^2-\vk^2-\mb^2]^{-1}$
are the Matsubara Green functions for the
free quark and the free scalar boson, 
respectively, and $\omega_n=(2n+1)\pi T$
and
$\nu_n=2n\pi T$ are the Matsubara frequencies for fermions
and bosons, respectively.
Taking the summation over the Matsubara frequency 
and making the analytic continuation 
$i\omega_n\rightarrow \omega+i\eta$,
one obtains the retarded self-energy 
$\Sigma^\mathrm{R}(\omega) 
= \tilde\Sigma(i\omega_n)|_{i\omega_n=\omega+i\eta}$.

The self-energy $\Sigma^\mathrm{R}(\omega) $ has 
an ultraviolet divergence which
originates from the $T$-independent part
$\Sigma_{T=0}^\mathrm{R}(\omega)\equiv \lim_{T\rightarrow
0^+}\Sigma^\mathrm{R}(\omega)$.
We remove this divergence by imposing 
the on-shell renormalization condition on
the $T$-independent part of the quark propagator;
for the details, see Appendix A.
The $T$-dependent part,
$\Sigma^\mathrm{R}_{T\not=0}(\omega)\equiv 
\Sigma^\mathrm{R}(\omega) - \Sigma^\mathrm{R}_{T=0}(\omega)$,
on the other hand, is free from divergence \cite{KKN2}.
For the numerical calculation,
it is convenient to first calculate 
$\im \Sigma^{\rm R}_{T\ne0}(\omega)$ 
and then determine the real part 
via the Kramers-Kronig relation
\begin{equation}
\re \Sigma^\mathrm{R}_{T\not=0}(\omega)=
\mathcal{P}\int_{-\infty}^{\infty}\frac{dz}{\pi}
\frac{\im \Sigma^\mathrm{R}_{T\not=0}(z)}{z-\omega},
\label{eq:resigpfinT}
\end{equation}
where $\mathcal{P}$ denotes the principal value.

The quark propagator for zero momentum is decomposed 
in the following way,
\begin{eqnarray}
G^{\rm R}(\omega)
&=&
G_+(\omega)\Lambda_+\gamma^0+
G_-(\omega)\Lambda_-\gamma^0,
\label{eq:G+-}
\end{eqnarray}
with projection operators 
$\Lambda_\pm \equiv ( 1\pm\gamma^0 )/2$ onto 
spinors whose chirality is equal($+$) or
opposite($-$) to the helicity.
We call $+$($-$)-sector as `quark'(`anti-quark') sector. 
Here
\begin{eqnarray}
G_\pm(\omega)
&=& \frac12 \Tr\left[G^{\rm R}(\omega)\gamma^0\Lambda_\pm\right]
\nonumber \\
&=& [ \omega + i\eta \mp m_f - \Sigma_\pm(\omega) ]^{-1},
\end{eqnarray}
with 
$\Sigma_\pm(\omega) 
= \rm{tr} [ \Sigma^R(\omega)\gamma^0 \Lambda_\pm]$.
From the charge conjugation symmetry,
we have a relation between $G_\pm$ \cite{LeBellac},
\begin{equation}
G_+ (\omega) = - G^*_- (-\omega).
\label{eq:GR+=GR-}
\end{equation}
Because of this relation, we consider only the `quark' sector $G_+(\omega)$.
For $m_f=0$, the propagator satisfies
$G_\pm (\omega) = -G^*_\pm (-\omega)$.
Notice that at finite quark number density,
the charge conjugation symmetry is violated and 
Eq.~(\ref{eq:GR+=GR-}) is no longer valid.

\begin{figure}[t]
\begin{center}
\includegraphics[clip,width=.4\textwidth]{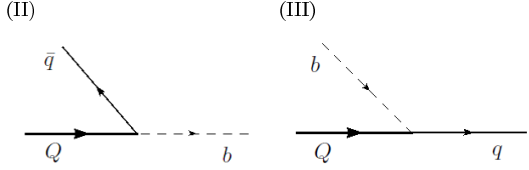}
\end{center}
\caption{
The decay processes corresponding to the terms (II) and (III)
in Eq.~(\ref{eq:imsigkin}).
}
\label{fig:landau}
\end{figure}

\begin{figure}[tn]
\includegraphics[clip,width=.49\textwidth]{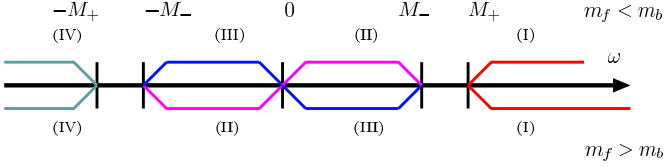}
\caption{The supports of the terms (I)-(IV) in Eq.~\eqref{eq:imsigkin}.
Here, $M_+ \equiv \mb+\mf$ and $M_- \equiv |\mb-\mf|$
are boundaries of the supports.
}
\label{fig:support}
\end{figure}

The imaginary part of $\Sigma_+(\omega)$ is given by
\begin{eqnarray}
\lefteqn{ \im\Sigma_+(\omega)
= -\frac{g^2}{2\pi}\int_{0}^\infty dk\frac{k^2}{E_k}\{ }
\nonumber \\
&& \phantom{+}\Tr[\Lambda_+\po{+}{\vk}] \delta(\omega-\ef-\eb)[1+n-f]
\mspace{14mu}\mathrm{(I)} \nonumber \\
&&+\Tr[\Lambda_+\po{-}{\vk}]\delta(\omega+\ef-\eb)[n+f]
\mspace{45mu}\mathrm{(II)} \nonumber \\
&&+\Tr[\Lambda_+\po{+}{\vk}]\delta(\omega-\ef+\eb)[n+f]
\mspace{45mu}\mathrm{(III)} \nonumber \\
&&+\Tr[\Lambda_+\po{-}{\vk}]\delta(\omega+\ef+\eb)[1+n-f]\}
\mspace{5mu}\mathrm{(IV)},
\nonumber \\
\label{eq:imsigkin}
\end{eqnarray} 
where
$\ef=[m_f^2+\vk^2]^{1/2}$, $\eb=[m_b^2+\vk^2]^{1/2}$,
$\po{\pm}{\vk}=(\ef\pm\gamma^0(\vec{\gamma}\cdot\vk+\mf))/(2\ef)$,
and $f = (\exp(\ef/T)+1)^{-1}$ and $n = (\exp(\eb/T)-1)^{-1}$
are the Fermi-Dirac and Bose-Einstein distribution functions, 
respectively.
Equation~(\ref{eq:imsigkin}) is 
proportional to the difference between the decay and creation
rates of the quasi-quark.
One can give a diagrammatical interpretation to each term (I)--(IV) 
\cite{BBS92,KKN2}.
In Fig.~\ref{fig:landau}, we show the decay processes
included in the terms (II) and (III), which are called the Landau damping.
The term (II) corresponds to 
a pair annihilation process 
between the quasi-quark ($Q$) and a thermally excited 
anti-quark ($\bar{q}$) with an emission of a boson ($b$),
and its inverse process.
The term (III) corresponds to
a scattering process of the quasi-quark by a thermally 
excited boson.
The Landau damping vanishes at $T=0$, as it 
involves thermally excited particles in the initial states.
As we will see later, the Landau damping 
is enhanced rapidly as $T$ goes high and plays a significant role
for the peak structure in the quark spectrum.

Each term (I)--(IV) has a support on the $\omega$ axis 
as shown in Fig.~\ref{fig:support}
owing to the energy-momentum conservation.
While any value of $\omega$ belongs to one support for  $m_f=0$,
the ranges $|m_b-m_f| < |\omega| < m_b+m_f$ are forbidden kinematically
for $m_f\ne0$ and hence $\im \Sigma_+(\omega)=0$ there.
In particular, the supports for the Landau damping (II) and (III) 
vanish for $m_f=m_b$.

The integral in Eq.~\eqref{eq:imsigkin}
can be performed analytically
to give
\begin{eqnarray}
\lefteqn{ \im \Sigma_+(\omega)
=-\frac{g^2}{64\pi} \frac{(\omega+\tm)(\omega-\dm)}{\omega^3} }
\nonumber \\
&&\times 
\sqrt{(\omega^2-\dm^2)(\omega^2-\tm^2)}
\nonumber \\
&&\times\left[\coth \frac{\omega^2+\dm\tm}{4\omega T}
+\tanh \frac{\omega^2-\dm\tm}{4\omega T} \right]
\nonumber \\
&&\times \left[\theta(\omega^2-\tm^2)-\theta(\dm^2-\omega^2)\right],
\label{eq:imsigplus}
\end{eqnarray}
where $\tm=\mf+\mb$, $\dm=|\mb-\mf|$, and 
$\theta(\omega)$ is the step function.

In the following sections,
we investigate the spectral properties of the quark
by analyzing the spectral function and  poles
of the quark propagator.
The spectral function is given by
$\rho(\omega) = -(1/\pi) \rm{Im} G^R(\omega)$
which is decomposed, similarly to Eq.~(\ref{eq:G+-}), as
$\rho(\omega) = \rho_+(\omega)\Lambda_+\gamma^0
+ \rho_-(\omega)\Lambda_-\gamma^0$,
with
\begin{eqnarray}
\rho_\pm(\omega)
&=&
-\frac1\pi \im G_\pm(\omega)
\nonumber \\
&=&
-\frac1\pi \frac{\im \Sigma_\pm(\omega)}
{(\omega\mp\mf-\re \Sigma_\pm(\omega))^2
+\im\Sigma_\pm(\omega)^2}.
\nonumber \\
\label{eq:rho_RI}
\end{eqnarray}
As directly derived from Eq.~\eqref{eq:GR+=GR-},
the charge conjugation symmetry leads to
\begin{equation}
\rho_+(\omega)=\rho_-(-\omega).
\end{equation} 
In general, $\rho_+(\omega)$ has several peaks
corresponding to the quasi-quark excitations.
Such excitations can also be expressed by complex poles 
of $G_+(\omega)$
which are given by solving
\begin{eqnarray}
[G_+(z)]^{-1} = z-\mf-\Sigma_+(z)=0,
\label{eq:pole}
\end{eqnarray}
with a complex number $z$.
The causality requires that the solutions of Eq.~(\ref{eq:pole}) 
are in the lower-half complex-energy plane.
The retarded self-energy $\Sigma_+(z)$ in Eq.~(\ref{eq:pole}) 
for a lower-half complex-energy is given by
\begin{align}
\Sigma_+(z) 
= \Sigma_+(\omega)|_{\omega\rightarrow z}
+ 2i \left.[\im \Sigma_+(\omega)\right]_{\omega \rightarrow z}.
\label{eq:G_+(z)}
\end{align}
Notice that the analytic continuation of the retarded function
to the lower-half plane requires the second term in 
Eq.~(\ref{eq:G_+(z)})
to compensate a cut of the first term on the real axis;
for details of this analytic continuation, see Ref.~\cite{KKKN}.
The residue of each pole reads
\begin{equation}
Z=\left[1-\frac{\partial \Sigma_+(z)}
{\partial z}\right]^{-1},
\label{eq:complexdispersion}
\end{equation}
at the position of the pole $z$. 
We remark that 
the residue of a complex pole is a complex number in general.

To understand how the peak structure 
in $\rho_+(\omega)$ emerges,
it is also convenient to use the notion of
``quasi-pole'' which is defined 
as zero of the {\em real} part of the inverse propagator;
\begin{equation}
\omega-\mf-{\rm Re}\Sigma_+(\omega)=0.
\label{eq:quasi-dispersion}
\end{equation}
Indeed, one sees from Eq.~(\ref{eq:rho_RI}) that
$\rho_+(\omega)$ becomes large at 
$\omega$ 
satisfying Eq.~(\ref{eq:quasi-dispersion}),
provided that ${\rm Im}\Sigma_+(\omega)$ is small there.

\section{The spectral functions of quarks 
coupled with a scalar boson }
\label{sec:spcsc}

In this section, we present the numerical results of 
the quark spectral function in the Yukawa model with a scalar boson.
In order to see the effect of the quark mass $m_f$,
we first recapitulate the result for $m_f=0$
\cite{KKN2},
focusing on its $T$-dependence at zero momentum.
We then show the numerical results for massive quarks.
The behavior of the self-energy is also analyzed for both cases.
We fix the coupling constant $g$ unity throughout this and 
subsequent sections.
We have checked that 
our results do not change qualitatively with the variation
of $g$ over a rather wide range \cite{KKN2}.

\subsection{Quark spectrum with $m_f=0$}
\label{ssc:spcscvmf}

\begin{figure*}[tn]
\begin{center}
\includegraphics[clip,width=.99\textwidth]{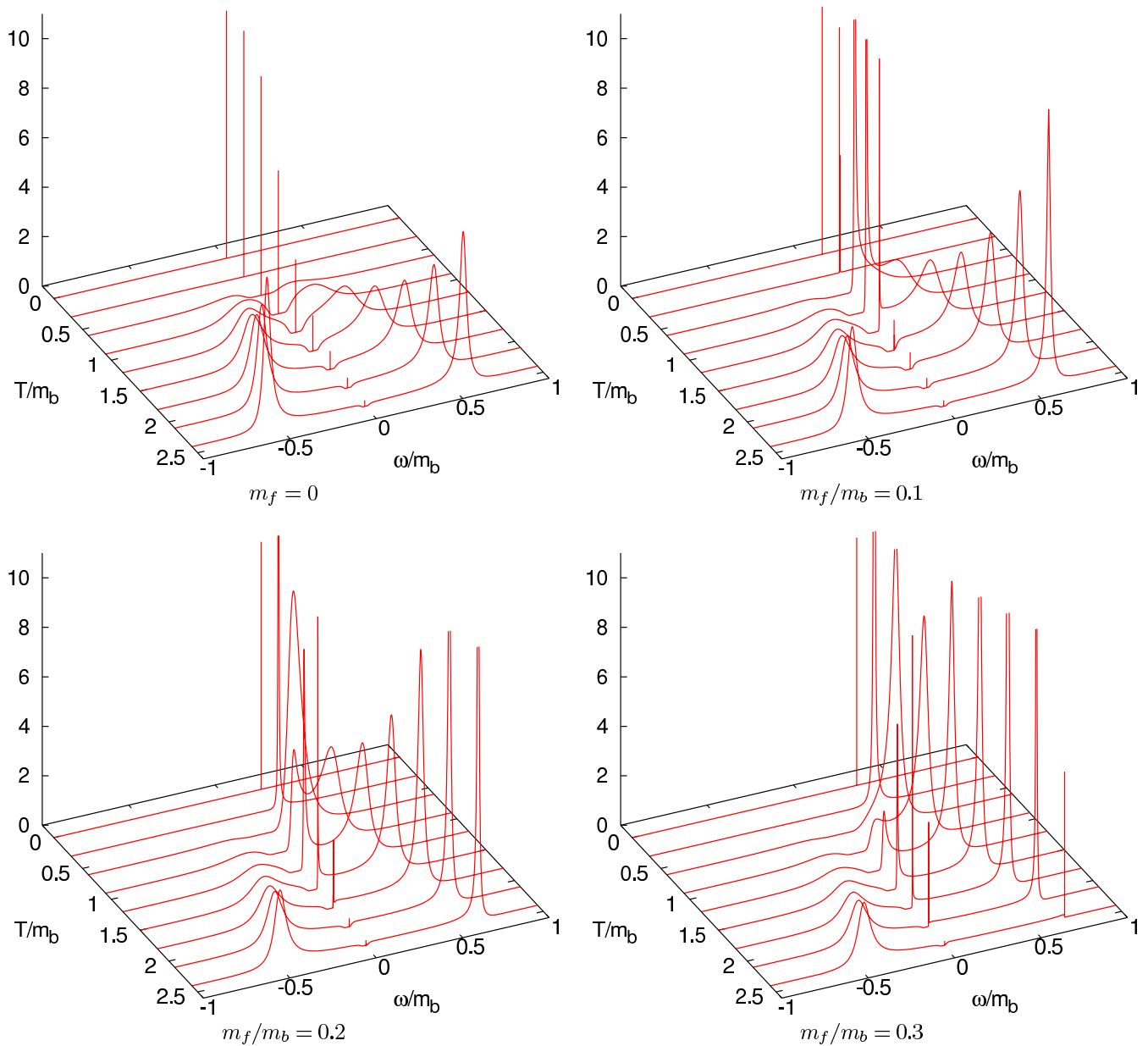}
\end{center}
\caption{The quark spectral functions $m_b \rho_+(\omega)$ 
for $\mf/\mb=0,0.1,0.2$ and $0.3$.
The heights of the peaks of the $\delta$-function shape 
represent $Z\times 10$
with $Z$ being the residue.
}
\label{fig:spc}
\end{figure*}

In the left upper panel of Fig.~\ref{fig:spc}, we show
the $T$-dependence of the spectral function $\rho_+(\omega)$
with $m_f=0$.
$\rho_+(\omega)$ 
takes on a peak with a $\delta$-function shape
at the origin independently of $T$.
In this figure, the height of this peak denotes $Z\times 10$ with $Z$
being the residue of the corresponding pole in the quark propagator.
At $T=0$, the residue is unity
owing to the on-shell renormalization condition.
As $T$ is raised, 
the magnitude of the present  residue decreases 
while two bumps newly  appear  at finite energies and
eventually turn to narrow peaks.
At intermediate temperatures, i.e.,  $\tau \equiv T/\mb = 1$-$1.5$,
$\rho_+(\omega)$ thus has a three-peak structure \cite{KKN1,KKN2}.
As will be shown in the next section, 
there exist poles corresponding to the new two peaks,
and the residues  of the three poles 
tend to have almost the same magnitude 
in  this range of  $T$,
which indicates that 
the three peaks are equally significant.
As $T$ is raised further, 
the residue of the pole at the origin damps quickly
and $\rho_+(\omega)$ is dominated by the two peaks 
at finite energies:
These two peaks can be identified 
with the normal quasi-quark and the anti-plasmino 
excitations \cite{KKN2}
which are familiar collective fermion excitations obtained 
in the HTL approximation in gauge theories \cite{LeBellac}.
The positions of these peaks correspond to the thermal masses.

\begin{figure}[tn]
\begin{center}
\includegraphics[clip,width=.49\textwidth]{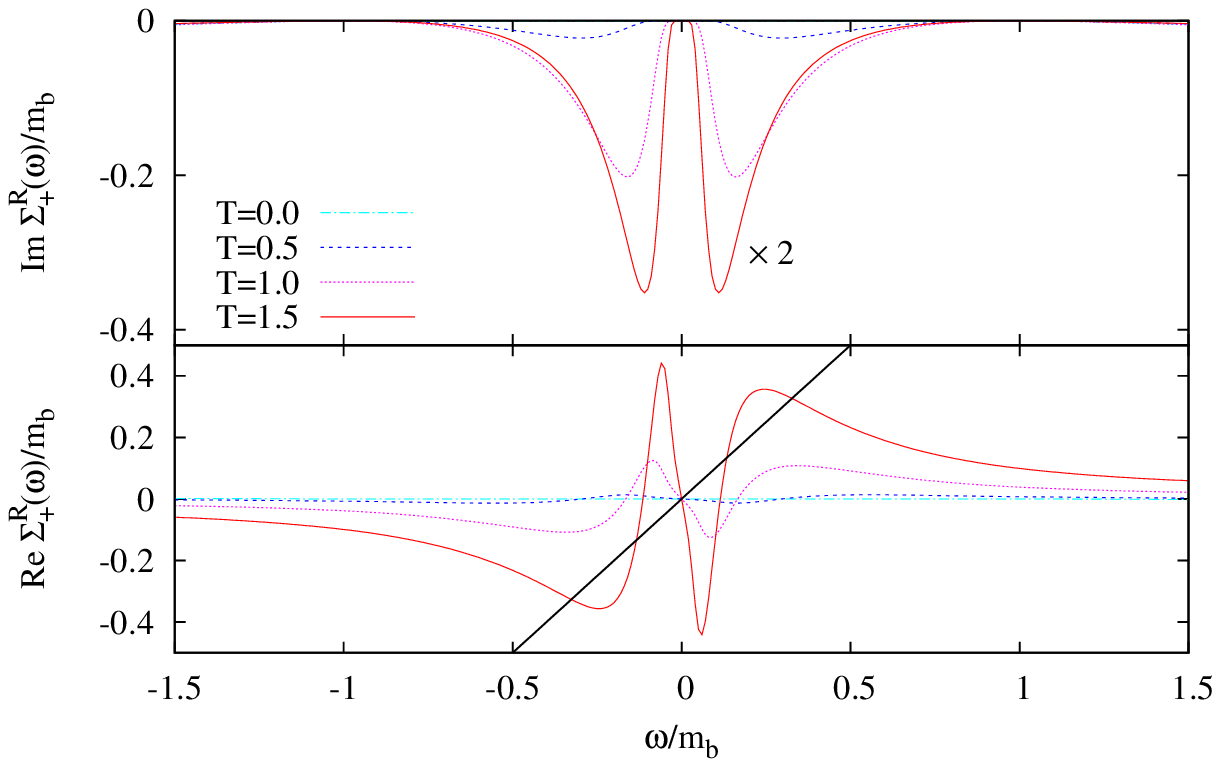}
\end{center}
\caption{
The quark self-energy $\Sigma_+(\omega)$ with $m_f=0$
for $T/\mb=0,0.5,1.0$ and $1.5$.
The upper and lower panels show the imaginary and real parts,
respectively.
}
\label{fig:se00}
\end{figure}

To understand the $T$-dependence of $\rho_+(\omega)$, 
we analyze  the imaginary and real parts of
$\Sigma_+(\omega)$, which are shown  in Fig.~\ref{fig:se00}.
For $m_f=0$, the self-energy $\Sigma_+(\omega)$ has 
symmetry properties, 
$\im\Sigma_+(\omega)=\im\Sigma_+(-\omega)$
and $\re\Sigma_+(\omega)=-\re\Sigma_+(-\omega)$.
At $T=0$, 
$\im \Sigma_+(\omega)$ takes non-zero values only for
$|\omega|>\mb$ corresponding to the supports of the
terms (I) and (IV) 
in Eq.~\eqref{eq:imsigkin}.
At $T\not= 0$, $\im\Sigma_+(\omega)$ gets to have 
a support in the region $|\omega|<m_b$ owing to the 
Landau damping, i.e. the terms (II) and (III),
and forms two peaks there. 
The heights of these peaks grow up quickly as $T$ is raised.

Then owing to  the Kramers-Kronig relation 
between the real and imaginary parts of the self-energy 
Eq.~\eqref{eq:resigpfinT},
$\re \Sigma_+(\omega)$
inevitably shows a steep rise in the two regions of $\omega$  
corresponding to the peak regions of $\im \Sigma_+(\omega)$ 
for $|\omega|<\mb$ \cite{KKN2}; 
their steepness and the magnitudes
are amplified as $T$ is raised.

This oscillatory behavior of $\re\Sigma_+(\omega)$ affects
the number of quasi-poles.
In the lower panel of Fig.~\ref{fig:se00}, we show a line 
$\re\Sigma_+(\omega)=\omega$
to find the solutions of Eq.~(\ref{eq:quasi-dispersion});
crossing points between this line and $\re\Sigma_+(\omega)$ 
give the quasi-poles.
The figure shows that at low $T$, there exists only 
one quasi-pole at the origin.
As $T$ is raised, the oscillatory behavior of $\re\Sigma_+(\omega)$ 
is so enhanced that  four new quasi-poles get to exist.
However, $\im\Sigma_+(\omega)$ has so large values 
at two of the quasi-poles that any peak structure in
$\rho_+(\omega)$ is  not formed at these points;
 thus we end up with
only the three peaks in the spectral function
at the origin and the other two quasi-poles 
where $\im\Sigma_+(\omega)$ is small.
The moral which one should learn from the above discussion
is that the Landau damping which causes two peaks in
$\im\Sigma_+(\omega)$ 
is essential for the formation 
of the three-peak structure in $\rho_+(\omega)$;
see Appendix~\ref{app:3peak} for an elucidating discussion 
with use of an analytic toy model for the self-energy.

\subsection{Quark spectral function with small quark mass}
\label{ssc:spcscfmf}

In this subsection, we investigate how a finite but small quark mass
affects and modifies the quark spectral function
$\rho_+(\omega)$.

We show $\rho_+(\omega)$ with $\mf/\mb=0.1$
in the upper-right panel of Fig.~\ref{fig:spc}.
At $T=0$, $\rho_+(\omega)$ takes on a peak with 
a $\delta$-function shape at $\omega=\mf$;
this pole always exists in $\rho_+(\omega)$ at $T=0$ 
irrespective of the value of $m_f$.
As $T$ is raised, this peak gets to have a width and
moves toward the origin.
There also appears a shoulder in the  positive-energy region larger
than $m_f$, 
which turns to a narrow peak at $\tau\equiv T/\mb\gtrsim 1.5$.
In the  negative-energy region, there appears also another peak
whose development is slower than that 
in the positive energy region.
The three-peak structure is then barely seen in $\rho_+(\omega)$ 
at $\tau\simeq1$.

In the lower-left panel of Fig.~\ref{fig:spc}, 
we show the spectral functions for $\mf/\mb=0.2$.
As mentioned above,
there exists a peak of the $\delta$-function type
at $\omega=m_f$ at $T=0$.
As $T$ is raised, this peak becomes broader and eventually splits 
into two peaks;
one of which approaches the origin while the other 
connects to the normal quasi-quark excitation having a non-zero
thermal mass at high $T$.
There also appears a broad bump in the negative-energy region,
which only gradually gets developed into a peak
at higher $T$ than that in the positive-energy region.
Thus, the number of clear peaks in $\rho_+(\omega)$ 
at a certain $T$
is two at most, and a clear three-peak structure is hardly seen.

In the lower-right panel of Fig.~\ref{fig:spc}, 
the spectral function $\rho_+$ for $\mf/\mb=0.3$ is shown.
We see that the peak at $\omega=m_f$ at $T=0$ 
smoothly connects to the normal quasi-quark excitation 
in the high $T$ limit, which is 
quite different
from the cases with $m_f/m_b\leq 0.2$.
Furthermore,
the three-peak structure in $\rho_+(\omega)$
does not appear for any $T$; this is because
the peak in the negative-energy region develops 
only very slowly with increasing $T$.
As we will see later in Sec.~\ref{sec:polesc},
this behavior 
reflects in the $T$-dependence of the
poles of the quark propagator.
This qualitative change of the $T$-dependence of the pole
behaviors can be understood in terms of the level crossing as 
will be discussed
in Sec.~\ref{sec:level-mixing}.

\begin{figure}[tn]
\begin{center}
\includegraphics[clip,width=.49\textwidth]{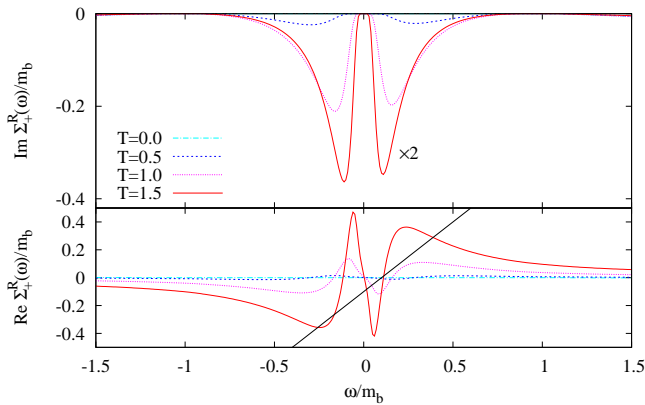}
\end{center}
\caption{
The quark self-energy $\Sigma_+(\omega)$ with $m_f/m_b=0.2$
for several values of $T$.
}
\label{fig:se01}
\end{figure}

The effects of $m_f$ on the behavior of $\rho_+(\omega)$ can 
be understood more precisely by analyzing the $m_f$-dependence of 
the self-energy and quasi-poles.
In Fig.~\ref{fig:se01}, we show $\Sigma_+(\omega)$ for $m_f/m_b=0.2$.
Comparing this figure with Fig.~\ref{fig:se00},
one finds that the 
behavior of $\Sigma_+(\omega)$ 
hardly changes with this slight change of the quark mass.
Since the quasi-poles are given by solving
$\re\Sigma_+(\omega) = \omega-\mf$, 
the dominant effect of $m_f$ thus comes from the quark mass term
in the r.h.s. of the above equation.
The line to determine the solutions of this equation 
is  shifted lower with $m_f\not=0$,
as shown in Fig.~\ref{fig:se01}.
Owing to this shift, there arises an asymmetry in the positions of 
quasi-poles in  positive- and negative-energy regions.
The two quasi-poles in the positive-energy region
emerge at lower temperature, while the quasi-pole
in the negative-energy region at higher temperature.

\section{The pole structure of quark coupled   
with scalar boson}
\label{sec:polesc}

In this section, we investigate the pole structure of 
the quark propagator $G_+(\omega)$.
We  find the three poles which correspond to 
the peaks in $\rho_+(\omega)$ discussed in the previous section,
and analyze their $T$-dependence.
It is found that the $T$-dependence of the poles
shows a drastic change at $m_f/m_b\simeq0.21$,
which means that the physics contents of the 
collective quark excitations are changed at this point.


\subsection{Pole structure of massless quark}
\label{ssc:polescvmf}

We first examine the case with $\mf=0$.
The poles of the quark propagator are given as complex roots
of Eq.~(\ref{eq:pole}).

We first notice that $G_+(z)$ has
an infinite number of poles in the lower-half complex-energy
plane.
In particular, there are two series of infinite poles
near the imaginary axis whose density increases infinitely
as they approach the origin.
The manifestation of these poles is mathematically attributed to 
the hyperbolic cotangent and tangent in Eq.~(\ref{eq:imsigplus}),
\begin{equation}
\coth\left(\frac{z^2+m_b^2}{4zT}\right) +
\tanh\left(\frac{z^2-m_b^2}{4zT}\right),
\end{equation}
where we have set $\mf=0$.
For $m_b \ne 0$,
the arguments of them diverge as $\sim1/z$ at the origin,
and these terms lead to a rapid oscillation of $\Sigma_+(z)$
in the complex-energy plane.
On the imaginary axis, for example, this term behaves as 
$\cot(1/\zeta) + \tan(1/\zeta)$ near the origin with $\zeta=-iz$, 
and thus $\im\Sigma_+(\zeta)$ has infinite number of divergences 
around the origin. This behavior also brings about singularities 
of $\re\Sigma_+(\zeta)$ through the Kramers-Kronig 
relation Eq.~(\ref{eq:resigpfinT}),
and leads to the infinite number of roots of Eq.~(\ref{eq:pole}) 
near the imaginary axis.
These poles, however, do not correspond to any peaks 
in the spectral function, and thus has no physical significance.
Because this oscillation comes from the $T$-dependent term,
this is specific at finite $T$.

Besides these poles, we have found that  there exist isolated three poles
which have a 
physical significance in the sense that they all correspond
to the peaks in $\rho_+(\omega)$.
One of them is that existing at the origin independently of $T$.
We label this pole as (A) in the following.
For $\tau\gtrsim 2.0$, we find the other
two poles associated with the sharp peaks in $\rho_+(\omega)$;
one  has a positive real part and is labeled as (B),
while the other has a  negative real part and
is labeled as (C).
As $T$ is lowered, the poles (B) and (C) move toward
the region containing the above-mentioned infinite poles and 
become hard to 
distinguish from them numerically:
For $\tau<0.4$, our numerical algorithm failed
to identify these two poles, and the pole search is stopped
at $\tau =0.4$.
In the following, we concentrate on the poles (A)--(C),
since they have  physical significance:
Indeed it will be shown that the peak structure
 of $G_+(\omega)$ is well 
described solely by these three poles.

\begin{figure}[tn]
\begin{center}
\includegraphics[clip,width=.49\textwidth]{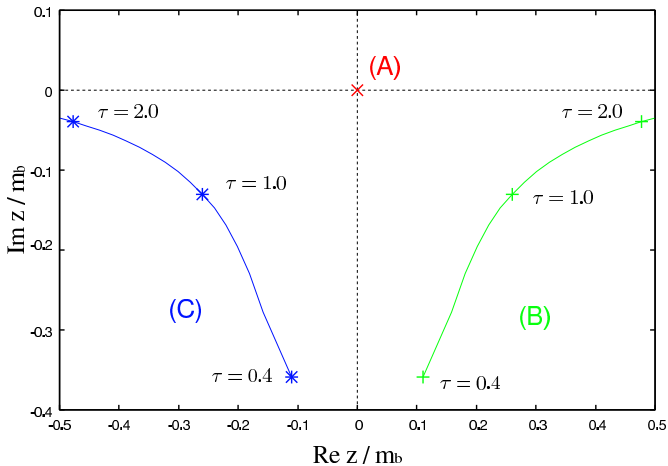}
\end{center}
\caption{
The $T$-dependence of the poles (A)--(C) with $m_f=0$.}
\label{fig:pole00}
\end{figure}

\begin{figure}[tn]
\begin{center}
\includegraphics[clip,width=.49\textwidth]{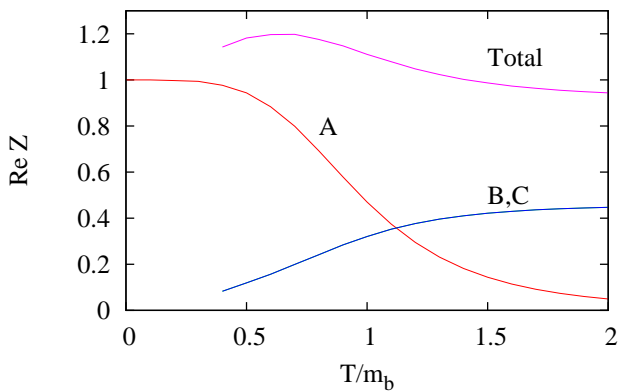}
\end{center}
\caption{
The $T$-dependence of ${\rm Re}Z$ of the poles (A)--(C) 
with $m_f=0$.
The summation of ${\rm Re}Z$, Eq.~(\ref{eq:sumBW}),
is also shown.
}
\label{fig:res00}
\end{figure}

\begin{figure*}
\begin{tabular}{ccc}
\includegraphics[clip,width=.33\textwidth]{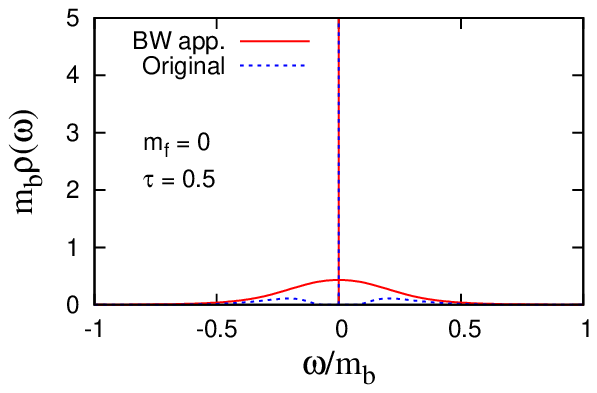}&
\includegraphics[clip,width=.33\textwidth]{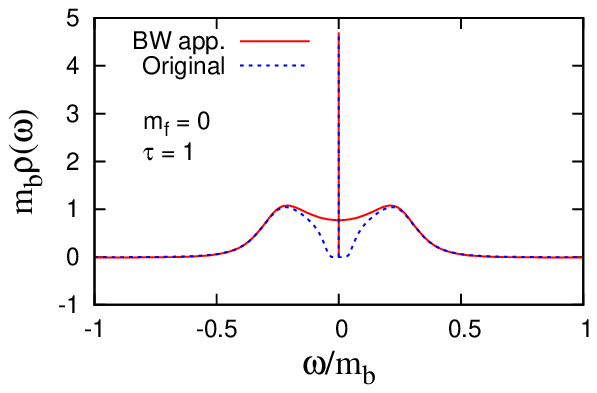}&
\includegraphics[clip,width=.33\textwidth]{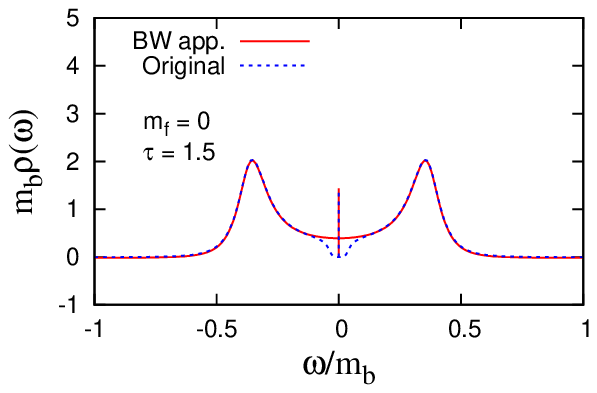}\\
\includegraphics[clip,width=.33\textwidth]{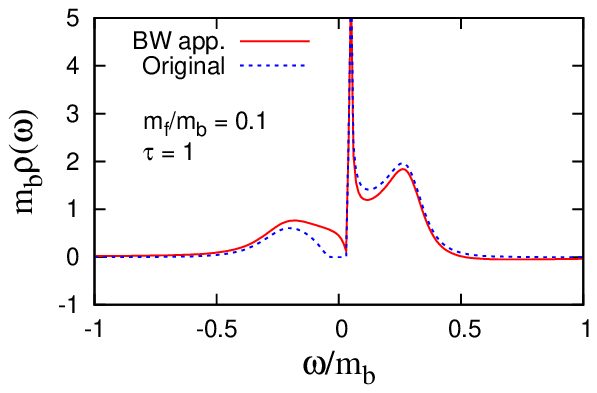}&
\includegraphics[clip,width=.33\textwidth]{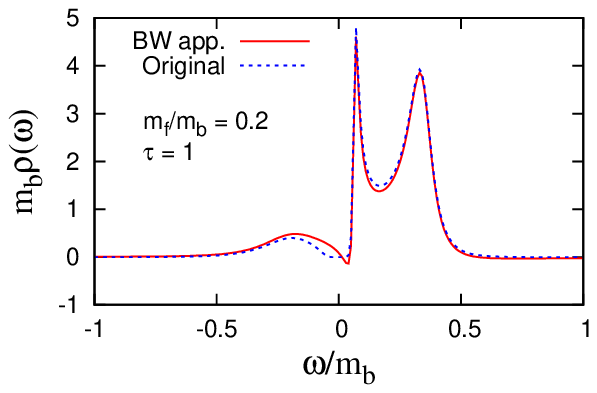}&
\includegraphics[clip,width=.33\textwidth]{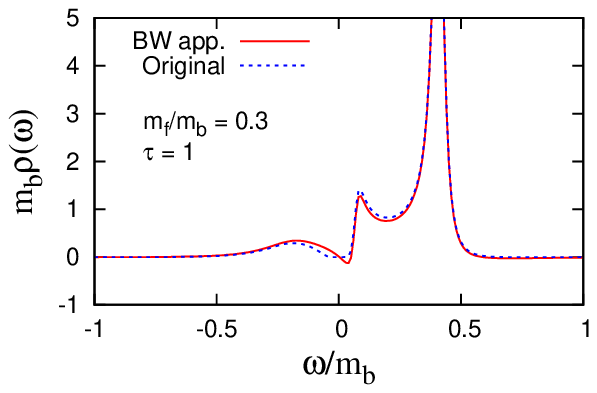}
\end{tabular}
\caption{
The comparison of the spectral functions with
the Breit-Wigner approximations $\rho^{\rm BW}_+(\omega)$ 
and the original one $\rho_+(\omega)$
for several quark masses and temperatures.
}
\label{fig:bw00}
\end{figure*}

In Fig.~\ref{fig:pole00}, we show the $T$-dependence of
the positions of the poles (A)--(C).
The real parts of the residues at these poles,
$\re Z_A$, $\re Z_B$ and $\re Z_C$,
are also shown in Fig.~\ref{fig:res00}.
One sees from these figures that for relatively low $T$, 
$|\im z_B|$ and $|\im z_C|$ are large and $\re Z_B$
and $\re Z_C$ are small, which
is consistent with the result in the previous section that
$\rho_+(\omega)$ is dominated by the pole (A) at 
sufficiently low $T$.
For $\tau\simeq1.2$, all the residues have the same value, which indicates
that the collective excitations corresponding
to these three poles have a similar strength.
For high temperatures, $|\im z_B|$ and $|\im z_C|$ 
become smaller in accordance with the formation of the sharp peaks 
in $\rho_+(\omega)$.

In order to check that the poles (A)--(C) certainly
correspond to the peaks in $\rho_+(\omega)$,
we compare the spectral function with that 
constructed from a Breit-Wigner (BW) approximation
\begin{equation}
\rho^\mathrm{BW}_+(\omega)
=-\frac{1}{\pi}\im \sum_{i=A,B,C}\frac{Z_i}{ \omega-z_i }.
\label{eq:BW}
\end{equation}
This approximated and the original spectral functions
for $\tau = 0.5,1$ and $1.5$
are plotted in the upper panels of Fig.~\ref{fig:bw00}.
One sees that $\rho^\mathrm{BW}_+(\omega)$
well reproduces the original spectral function $\rho_+(\omega)$, 
especially for higher $T$.
We thus conclude that the three poles well represent 
the excitation property of the quasi-quark.
The BW approximation, however, fails to reproduce the precise
shape of $\rho_+(\omega)$
around the origin,
which is due to the contributions of 
the infinite series of poles near the imaginary axis.

The fermion spectral function satisfies the sum rule
\begin{equation}
\int_{-\infty}^\infty d\omega\rho_+(\omega)=1.
\label{eq:sum-rule1}
\end{equation}
If the poles (A)--(C) well describe the whole 
analytic structure of the propagator,
the sum of the residues $\sum_i Z_i$ should take a value
close to unity, provided that 
a possible violation of the sum rule
owing to a renormalization procedure is negligible
\cite{LeBellac}.

Here, we first notice that, with the BW approximation Eq.~(\ref{eq:BW}),
the left hand side of Eq.~\eqref{eq:sum-rule1} is given by,
\begin{equation}
\int_{-\infty}^\infty d\omega \rho^\mathrm{BW}_+(\omega)
=\sum_{i=A, B, C}\re Z_i\equiv Z_{\rm Tot},
\label{eq:sumBW}
\end{equation}
which tells us that only the real part of the residue contributes
to the strength of the spectrum.
 The $T$ dependence of
 $Z_{\rm Tot}$ as well as $\re Z_i$ ($i=$A, B and C) are shown
in Fig.~\ref{fig:res00}.
One sees that 
the sum $Z_{\rm Tot}$ gives a value close to unity
in the whole range of $T$
with a small deviation,  about $20\%$ at most.
We remark that if the values of residues of the infinite series of poles
around the imaginary axis is added to $Z_{\rm Tot}$
the deviation from unity becomes much smaller.

\begin{figure}[tn]
\includegraphics[clip,width=.49\textwidth]{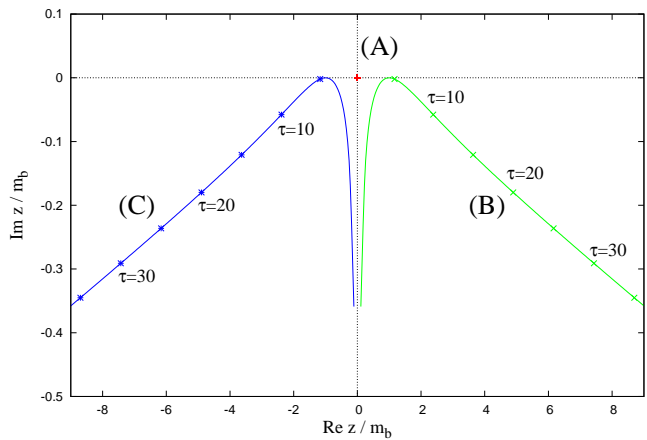}
\caption{
The $T$-dependence of the poles (A)--(C) with $m_f=0$
up to
$\tau\simeq40$.
The points denote the position of poles 
for $\tau=5,10,15,\cdots$.}
\label{fig:htlpole}
\end{figure}

Finally, we examine the poles (A)--(C) for $\tau \gg 1$.
In Fig.~\ref{fig:htlpole}, we show the poles (A)--(C)
up to $\tau\simeq40$.
The poles (B) and (C) reach the real axis at $\omega=m_b$
when $\tau\simeq 4.2$. 
This is due to a suppression of the phase space of 
the decay processes at $\omega=m_b$.
For higher $T$, $|\im z/m_b|$ for the poles (B) and (C)
increase monotonically with $T$.
At sufficiently high $T$, 
the real parts of the poles (B) and (C) take a value close to
those determined in the HTL approximation having
the thermal mass $m_T=gT/4$ \cite{KKN2},
although there is a small deviation owing to the strong coupling $g=1$
taken here.
The behavior of the imaginary part of each pole is also consistent 
with that in the HTL approximation which gives the width of order $g m_T$
\cite{LeBellac}.
Although there always exists the pole (A) at the origin,
the residue of this pole decreases 
and becomes negligible in the high-$T$ limit.

\subsection{Pole structure of massive quark}
\label{ssc:polescfmf}

In this subsection, we examine the $T$-dependence
of the pole structure of the propagator
for the  massive quarks by varying the quark mass 
($\mf/\mb=0.1 \rightarrow 0.3$).
We shall show how the small but non-vanishing
quark mass significantly affects the physics contents of the quasi-quark
excitations at finite $T$.

First of all, we have found that 
there exist the three poles of physical significance
which correspond to a peak or bump in the spectral function 
even for the massive quarks, too:
The lower panels of Fig.~\ref{fig:bw00}, 
 show the spectral functions in the BW approximation
$\rho^{\rm BW}_+(\omega)$ 
for $m_f/m_b=0.1,0.2$ and $0.3$ with fixed $\tau=1$.
We see that the BW approximation solely 
with the three poles well reproduces
the original spectral functions even for the cases of 
massive quarks, which means that
the three poles  well represent the physical excitations 
of quarks even with finite $m_f$.
Here we mention that an infinite number of poles due to a kinematical origin 
exist also for finite $m_f$
around the imaginary axis as in the case of $m_f=0$.

\begin{figure}[tn]
\includegraphics[clip,width=.49\textwidth]{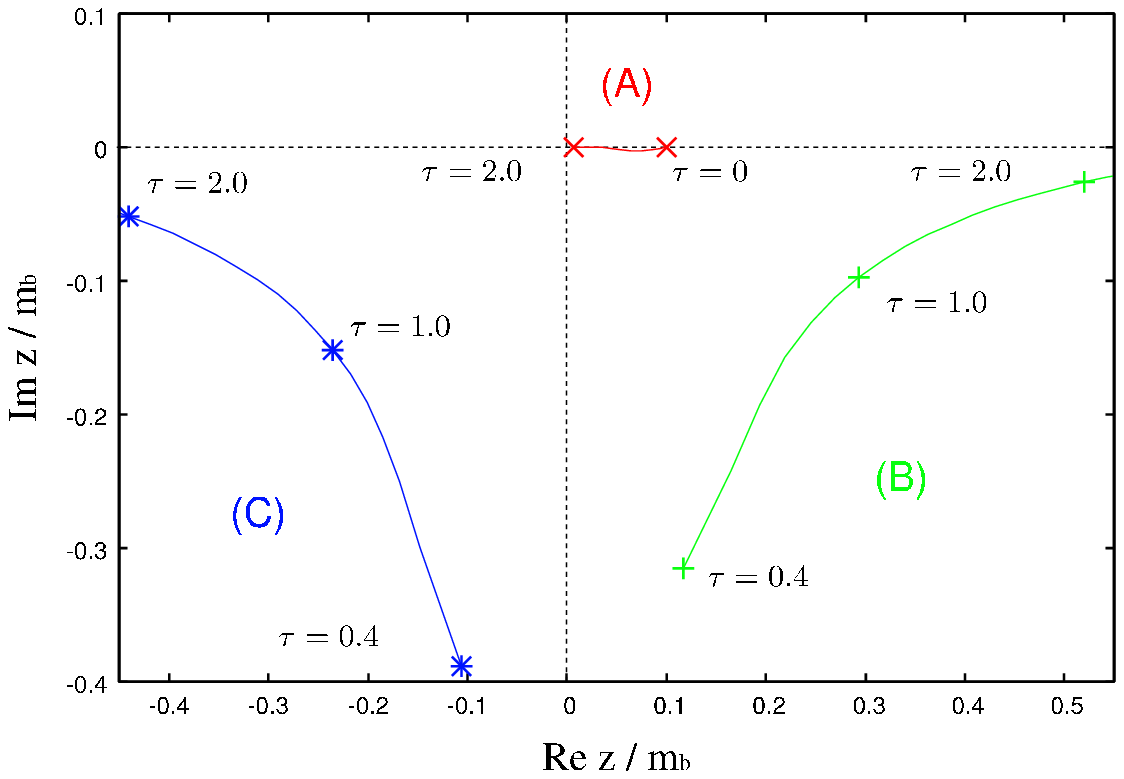}
\caption{
The $T$-dependence of the poles (A)$\sim$(C)
with $m_f=0.1$.
}
\label{fig:pole01}
\end{figure}

\begin{figure}[tn]
\begin{center}
\includegraphics[clip,width=.49\textwidth]{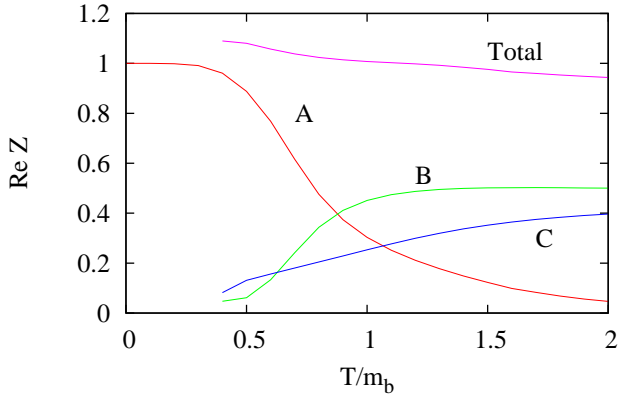}
\end{center}
\caption{
The $T$-dependence of ${\rm Re}Z$ with $m_f=0.1$.
}
\label{fig:res01}
\end{figure}

We show the $T$ dependence of the poles for $\mf/\mb=0.1$
in Fig.~\ref{fig:pole01}.
The figure  shows that 
the pole (A) moves toward the origin as $T$ is raised,
while the poles (B) and (C) move toward the real axis almost symmetrically 
with respect to the imaginary axis:
The small asymmetry of the two trajectories 
is caused by the fact that 
$|{\rm Im}z_B|$ is always smaller than $|{\rm Im}z_C|$
in the range of $\tau$ shown in the figure.
This inequality for the imaginary parts
is consistent with the behavior of $\rho_+(\omega)$
that the peak in the positive-energy region is sharper than 
that in the negative-energy one.
In Fig.~\ref{fig:res01},
we show $\re Z_i$ of each pole in the case of $m_f/m_b=0.1$.
The $T$-dependence of each residue is qualitatively the same
as that for $m_f=0$.
We notice that the residues for all the poles
have similar values at $\tau\simeq 1$
where a three-peak structure is barely seen in
$\rho_+(\omega)$.
For temperatures much higher than those shown in Fig.~\ref{fig:pole01},
the poles (B) and (C) eventually reach the real axis
at $|\re z_{B,C}| = m_b-m_f$.
As $T$ rises further, they move on the real axis 
until $|\re z_{B,C}| = m_b+m_f$ at which the poles start to leave 
the real axis. 
This movement of the poles along the 
real axis for some temperatures
reflects the support structure of the 
${\rm Im}\Sigma_{+}(\omega)$ as shown
in Fig.~\ref{fig:support}.
For $\tau\gg1$, the positions of the poles approach
those for $m_f=0$ shown in Fig.~\ref{fig:htlpole}.

\begin{figure}[tn]
\begin{center}
\includegraphics[clip,width=.49\textwidth]{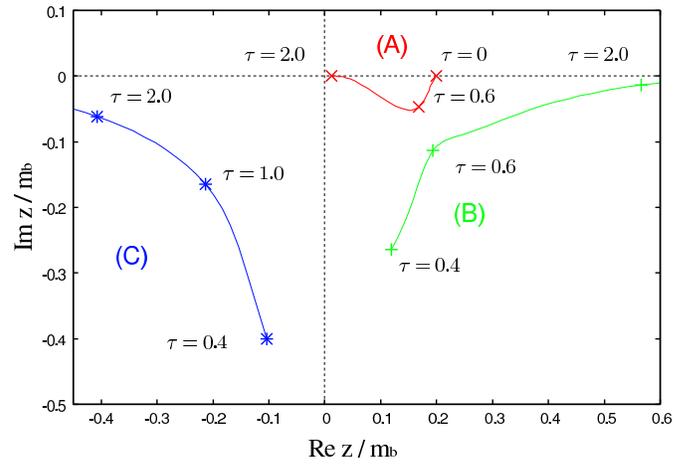}
\end{center}
\caption{
The $T$-dependence of the poles (A)--(C) with $m_f/m_b=0.2$.
}
\label{fig:pole02}
\end{figure}

\begin{figure}[tn]
\begin{center}
\includegraphics[clip,width=.49\textwidth]{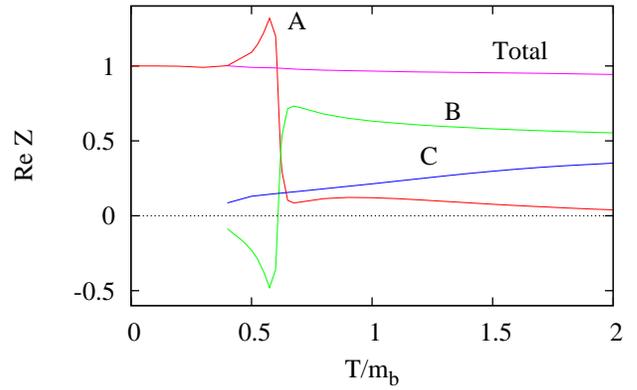}
\end{center}
\caption{
The $T$-dependence of ${\rm Re}Z$ with $m_f/m_b=0.2$.
}
\label{fig:res02}
\end{figure}

Let us turn to the case with $\mf/\mb=0.2$.
The $T$-dependence of the poles in this case is shown 
in Figs.~\ref{fig:pole02}:
Although the topology of the set of the trajectories and the behaviors 
of the poles 
along with the variation of $T$ are  qualitatively the same as before,
a notable point is that the trajectories of the poles (A) and (B) 
are so deformed that they tend to get close 
to each other around $\tau=0.6$. As we will see later,
this is actually a 
precursory deformation leading to a structural rearrangement of the
trajectories to be seen for the larger $m_f/m_b$.

The $T$-dependence of the pole residues $\re Z_i$
 ($i=$ A, B, C) is shown in Fig.~\ref{fig:res02}.
We see that the behavior of them 
are quite different from those for  smaller quark masses:
As $T$ is raised,
$\re Z_A$ first increases and then rapidly drops
at $\tau \simeq 0.6$, while $\re Z_B$ behaves oppositely.
The sum of the residues
$Z_{\rm Tot}$ is, however, 
almost unity for any $T$.
It should be  noted that although $\re Z_B$ is negative around $\tau=0.6$,
the pole (B) does not correspond to any peak in the spectral function 
 nor hence 
describe a physical excitation.

\begin{figure}[tn]
\begin{center}
\includegraphics[clip,width=.49\textwidth]{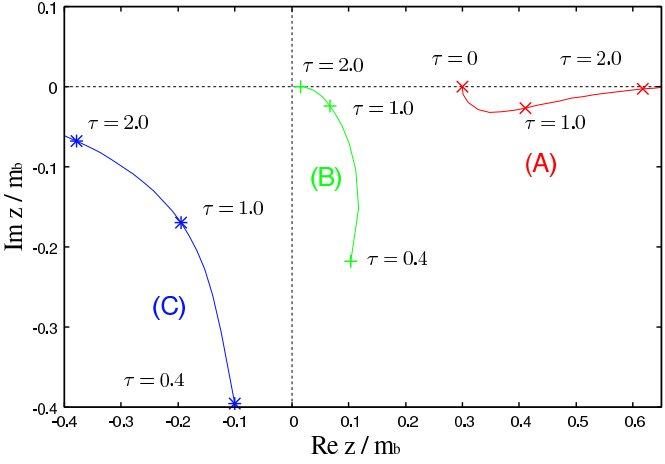}
\end{center}
\caption{
The $T$-dependence of the poles (A)--(C)
with $m_f/m_b=0.3$.
}
\label{fig:pole03}
\end{figure}

\begin{figure}[tn]
\begin{center}
\includegraphics[clip,width=.49\textwidth]{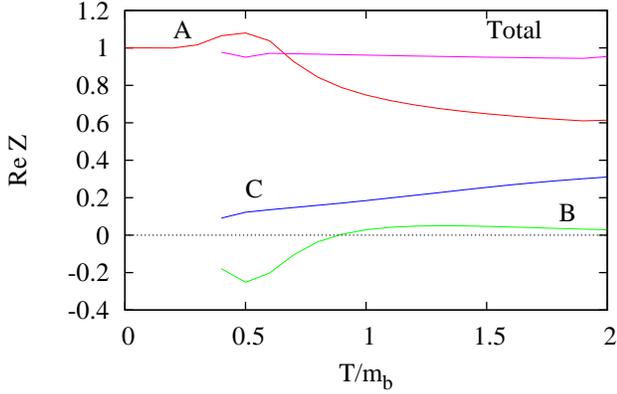}
\end{center}
\caption{
The $T$-dependence of ${\rm Re}Z$ of the poles (A)--(C) with 
$m_f/m_b=0.3$.
}
\label{fig:res03}
\end{figure}

Finally let us examine the case 
with $\mf/\mb=0.3$. 
 Fig.~\ref{fig:pole03} shows how the three poles
move along with the increase of  $T$.
At a first glance, one can see that 
a drastic change has occurred by a small increase of $m_f/m_b$ from 0.2
in the topology of the set of the three trajectories as well
as in the $T$-dependence of the poles.
The pole (A), which  is identified 
 as one  smoothly connecting 
to  $\omega=m_f$ at $T=0$, 
moves with increasing $T$ toward large positive energies
with a small or vanishing imaginary part, 
as pole (B) did in the previous cases,
 instead of going to the origin as the pole (A) did before.
On the other hand, the pole (B) 
first appears with a large imaginary part near the 
imaginary axis at  $\tau=0.4$ and tends to move toward
the real axis at high temperatures
as in the cases of $m_f/m_b\leq 0.2$, 
but then changes the direction and approaches the origin 
as $T$ is raised further, as the pole (A) did in the previous cases.
We now see that 
the poles (A) and (B)  at high temperatures
has exchanged their characters with each other; this 
exchange should have occurred 
at a critical quark mass $m_{fc}$ which is located between 
$0.2\mb$ and $0.3\mb$. The character exchange of the poles
is a kind of level crossing familiar in Quantum mechanics. 
In geometrical terms, 
when $m_f=m_{fc}$, the trajectories of pole (A) and (B) 
are so deformed that the two trajectories
would touch each other at some temperature
($\tau \sim 0.6$), and then for larger quark masses than 
$m_{fc}$,
the two trajectories
would be rearranged to form trajectories as shown 
in  Fig.~\ref{fig:pole03}. 
Indeed, we have numerically confirmed that the trajectories of 
the poles (A) and (B) get close to and then
touch each other
eventually at $m_f/m_b\simeq0.21$, i.e., $m_{fc}\simeq 0.2 \mb$.
The $T$-dependence of $\re Z$ of each pole for $\mf/\mb=0.3$,
shown in Fig.~\ref{fig:res03},
can also be understood naturally with the notion of the level
crossing.

\section{Spectral properties of quarks 
coupled with pseudoscalar bosons}
\label{sec:PS}

In this section,
 we replace the scalar boson with a pseudoscalar one and 
investigate how the behavior of the spectral function
and the pole structure change.
Throughout this section, 
we put a superscript ``PS'' to all the functions
where PS is for pseudoscalar boson.
When we refer to the functions calculated in the previous 
sections, we put a superscript ``S'' where
S is for scalar boson.

The Lagrangian density of the Yukawa model 
composed of  a quark and a pseudoscalar boson is given by,
\begin{equation}
\mathcal{L_\mathrm{PS}}
=\bar{\psi}(i\Slash{\partial} - \mf)\psi
+\frac{1}{2}\left[(\partial_\mu\phi)^2-\mb^2\right]
-g\bar{\psi}(i\gamma_5\phi)\psi.
\end{equation}
The quark self-energy in this model in the imaginary time
formalism at one loop reads
\begin{multline}
\tilde{\Sigma}^\mathrm{PS}(i\omega_m)
=-g^2T\sum_n\int\frac{d^3k}{(2\pi)^3}
i\gamma_5\mathcal{G}_0(\vect{k},i\omega_n)\\
\times i\gamma_5\mathcal{D}(-\vect{k},i\omega_m-i\omega_n).
\label{eq:sigma_ps}
\end{multline}
The difference between Eqs.~(\ref{eq:sigma_ps}) and 
(\ref{eq:sigma}) is the two factors of $i\gamma_5$ only.
As one can easily check,
this difference leads to the following simple relations
for the retarded self-energies,
\begin{eqnarray}
{\rm Re}\Sigma^\mathrm{PS}_+(\omega)
&=& -{\rm Re}\Sigma^\mathrm{S}_+(-\omega),
\\
{\rm Im}\Sigma^\mathrm{PS}_+(\omega)
&=& {\rm Im}\Sigma^\mathrm{S}_+(-\omega).
\label{eq:symmetry}
\end{eqnarray}
For $\mf=0$, one immediately obtains $\Sigma^\mathrm{PS}_+(\omega)
=\Sigma^\mathrm{S}_+(\omega)$, and thereby
the type of bosons does not affect the quark propagator at all
\cite{KKN2} in accordance with
the chiral symmetry.

\begin{figure}[tn]
\includegraphics[clip,width=.49\textwidth]{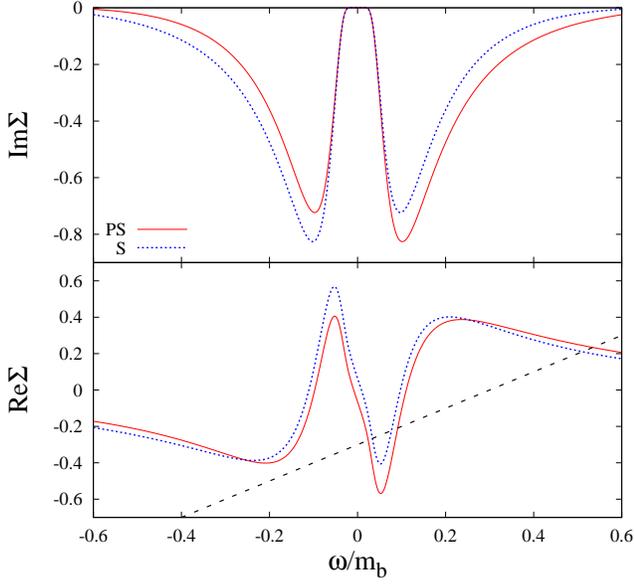}
\caption{The imaginary and real parts of the self-energies
$\Sigma^\mathrm{S}(\omega)$ and $\Sigma^\mathrm{PS}(\omega)$
for $\mf/\mb=0.3$ at $\tau=1.5$.
The dashed line is the line $\omega-\mf$.}
\label{fig:compse03}
\end{figure}

In Fig.~\ref{fig:compse03},
we show the imaginary and real parts of the self-energy
with the scalar and the pseudoscalar bosons
for $\mf/\mb=0.3$ and $\tau=1.5$.
We see that the behavior of them is qualitatively the same
as that for the scalar boson.

\begin{figure}[tn]
\includegraphics[clip,width=0.49\textwidth]{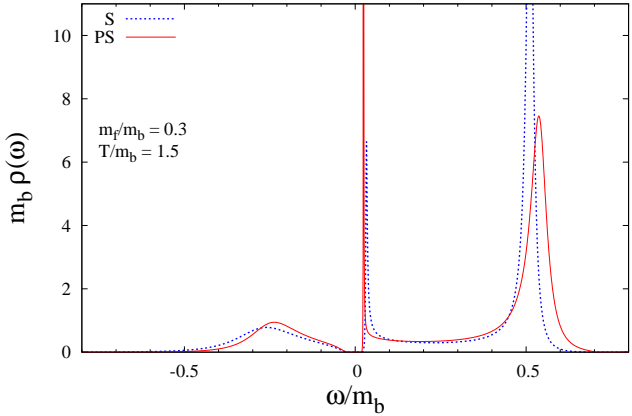}
\caption{The comparison of the spectral functions of
the quark coupled with the scalar and pseudoscalar bosons.
The labels ``S'' and ``PS'' stand for the scalar and 
pseudoscalar bosons, respectively.}
\label{fig:com03}
\end{figure}

\begin{figure}[tn]
\includegraphics[clip,width=.49\textwidth]{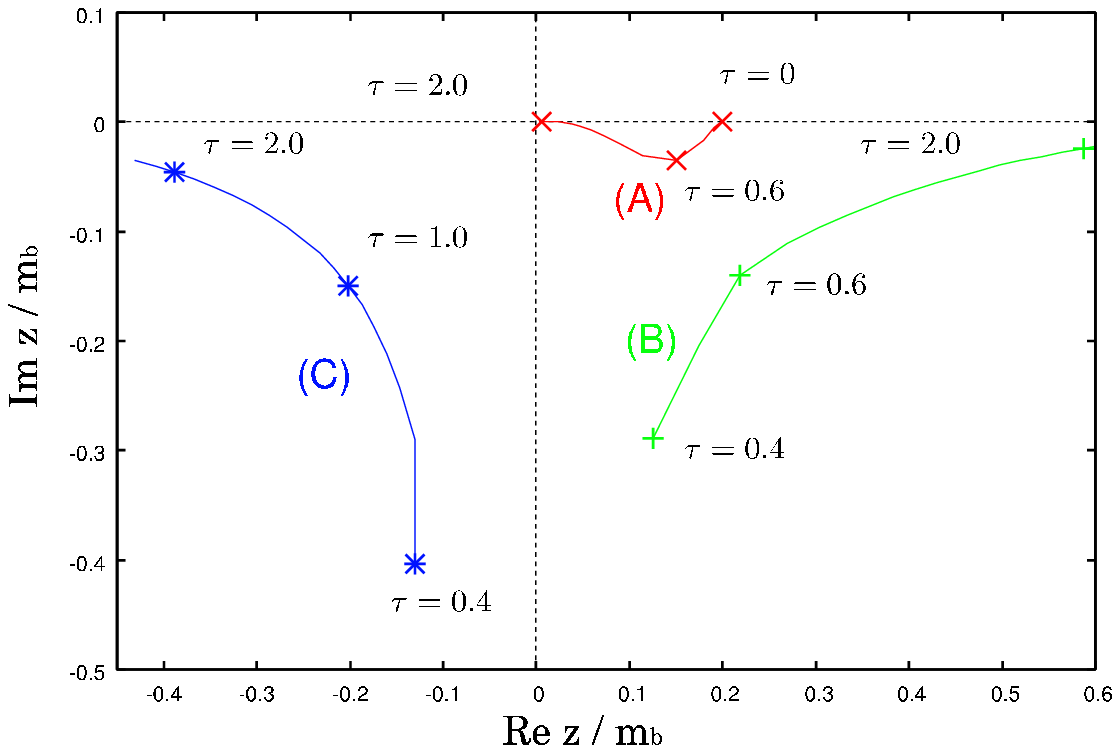}
\includegraphics[clip,width=.49\textwidth]{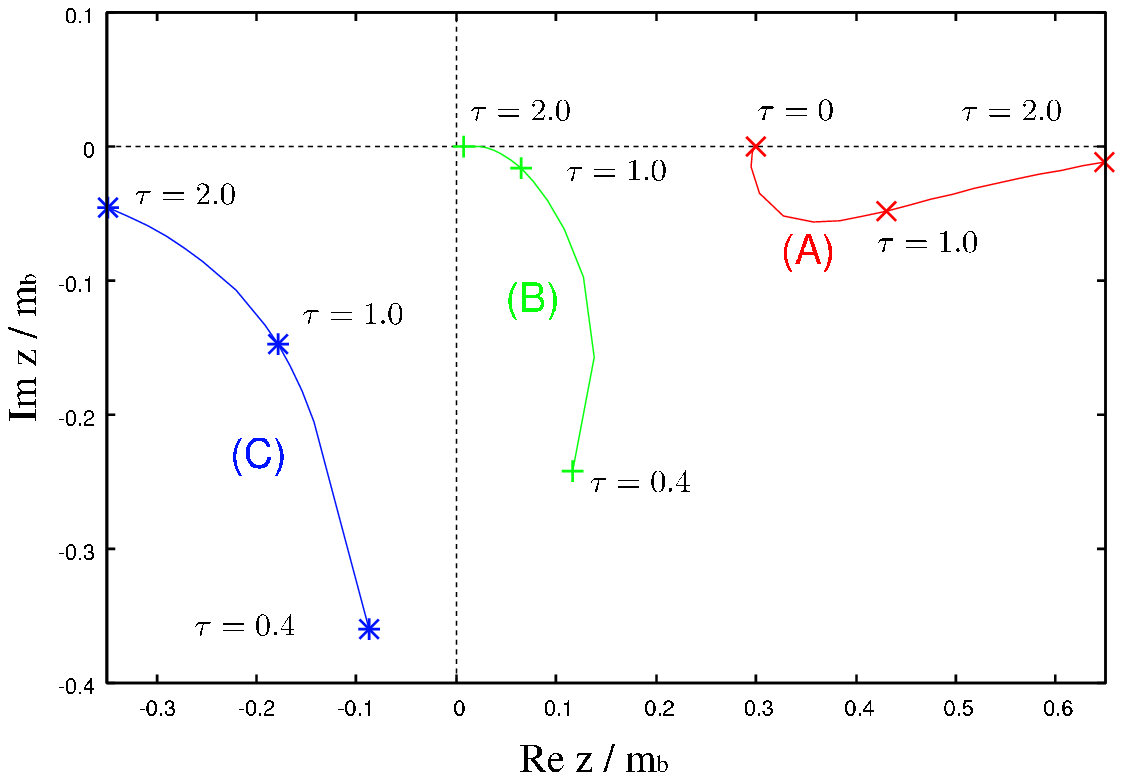}
\caption{The $T$-dependence of the poles.
The upper and lower panels are 
for $\mf/\mb=0.2$ and $0.3$, respectively.}
\label{fig:plps}
\end{figure}

From this feature, one naturally expects that 
the resulting spectral function and the pole structure
are also insensitive to the type of the boson.
In Fig.~\ref{fig:com03}, we show 
the spectral functions $\rho^\mathrm{PS}_+(\omega)$ 
and $\rho^\mathrm{S}_+(\omega)$
for $\mf/\mb=0.3$ and $\tau=1.5$.
One sees that the peak structures of these spectral functions
are qualitatively the same, indeed.
The pole position is shown 
in Fig.~\ref{fig:plps} for $\mf/\mb=0.2$ and $0.3$.
We finds again that each pole behaves similarly as before.
Quantitatively, $|\im z_A|$ is always smaller than that
obtained in the previous section,
while $|\im z_B|$ is larger.
The level crossing between the poles (A) and (B) occurs for this 
case also at $m_f/m_b\simeq0.23$, 
which is slightly larger than that for the scalar boson.

\section{Discussions; level mixing induced by resonant scattering}
\label{sec:level-mixing}

In the preceding sections, we have studied the spectral properties 
of quarks coupled with scalar and pseudoscalar bosons by examining 
the spectral function $\rho_+(\omega)$ and the pole structure of the 
quark propagator.
We have found that, although the three-peak structure for $m_f=0$ survives
even for finite $m_f$, such structure gradually ceases to exist and 
the number of clear poles in $\rho_+(\omega)$ becomes two at most for 
$m_f/m_b\gtrsim0.2$.
The behavior of $\rho_+(\omega)$ and the position of poles at finite 
$m_f$ show a drastic change as if there is a level crossing at 
$m_f/m_b\simeq0.21$ ($0.23$) when coupled with a scalar (pseudoscalar) 
boson.
In this section, we elucidate the mechanism for realizing this behavior 
in terms of the notion of the so called resonant scattering
\cite{rscatt}. 

We first recall that 
the formation of two peaks in $\im \Sigma^\mathrm{R}_+(\omega)$
plays a decisive role to cause
the three-peak structure in $\rho_+(\omega)$ 
(See, Sec.~\ref{sec:spcsc} and App.~\ref{app:3peak}.).
The decay rates corresponding to these two peaks 
are due to the Landau damping, which is
diagrammatically depicted in Fig.~\ref{fig:landau}:
The term (II) is the pair annihilation process 
of the quasi-quark ($Q$) and a thermally excited anti-quark ($\bar q$) 
into a boson $b$, $Q + \bar{q}\rightarrow b$, and its inverse process. 
Here, we notice that 
the annihilation of a thermally excited anti-quark can be regarded
as the creation of a hole with a positive quark number 
in the anti-quark distribution.
With this interpretation, the term (II) is
schematically described as $Q \to \bar{q}^{\rm hole} + b$;
the quasi-quark is converted to a hole in the anti-quark distribution
with an emission of a boson.
The term (III) is schematically depicted as $Q + b\to q$.
This process has a support for $\omega<0$ 
and the quasi-quark $Q$ is dominated by hole components 
of anti-quarks in this energy range.
Through the process of the term (III), the quasi-quark is 
converted to the on-shell quark, and thus 
the mixing between a quark and a hole of anti-quark is induced.

Scattering processes of a boson would induce a mixing between particle
and hole states at finite $T$, which are called the resonant scattering 
\cite{rscatt,KKN1,KKN2}.
The resonant scattering causes
a level repulsion in the energy spectra of the fermionic excitations.
In the case of superconducting phase transition, for example,
the preformed-pair modes induce a 
resonant scattering between particle and hole states
in the Fermi sea; the outcome is 
a gap-like structure around the Fermi surface, i.e., 
the so called {\em pseudogap},
 in the fermion spectral function \cite{rscatt}.
For the chiral phase transition,
it was shown \cite{KKN1,KKN2} that 
a resonant scattering 
between a quark and an anti-quark hole is induced through 
the coupling with the chiral soft modes;
there appears a three-peak structure in the quark spectral function 
$\rho_+(\omega)$ in the  low-energy and low-momentum region
\cite{KKN1},
see, Eqs.~(3.18) and (3.19) in Ref.~\cite{KKN2}.

\begin{figure}[tn]
\begin{center}
\begin{tabular}{cc}
\includegraphics[clip,width=.23\textwidth]{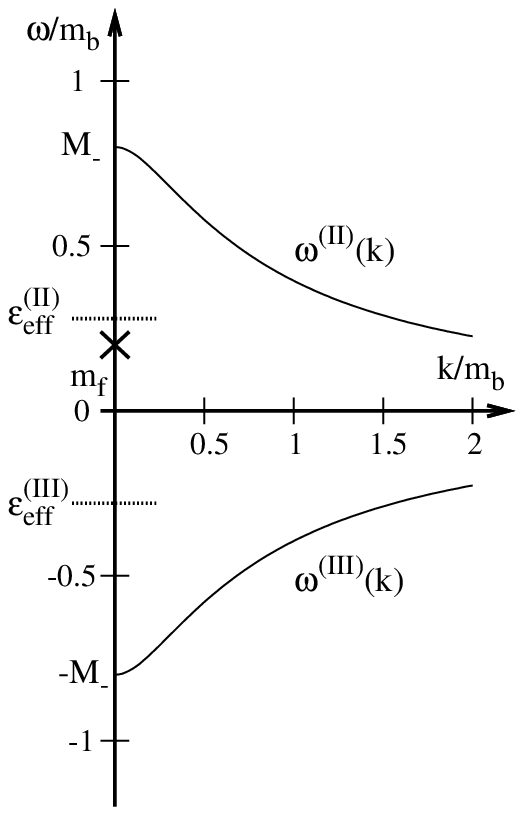}&
\includegraphics[clip,width=.23\textwidth]{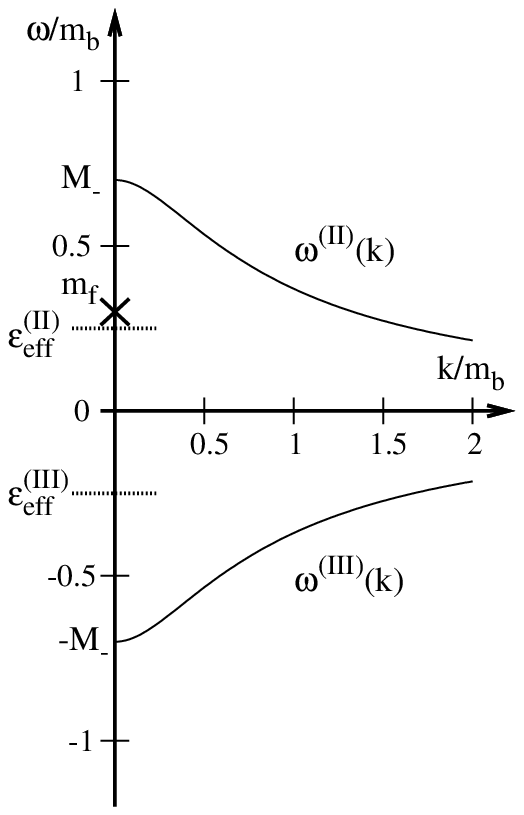}
\end{tabular}
\end{center}
\caption{
The energy levels for $m_f/m_b=0.2$ (left) and 0.3 (right).
The solid lines are $\omega^{\rm (II)}(k)$ for $\omega>0$
and $\omega^{\rm (III)}(k)$ for $\omega<0$.
The crosses indicate the free-quark energy at zero momentum,
$\omega=m_f$.
The dashed lines $\varepsilon_{\rm eff}^{\rm (II)}$ and
$\varepsilon_{\rm eff}^{\rm (III)}$
denote the most probable energy levels
of $\omega^{\rm (II)}(k)$ and $\omega^{\rm (III)}(k)$,
respectively.. 
}
\label{fig:level-mixing}
\end{figure}

\begin{figure*}[tn]
\begin{tabular}{ccc}
\includegraphics[clip,width=.32\textwidth]{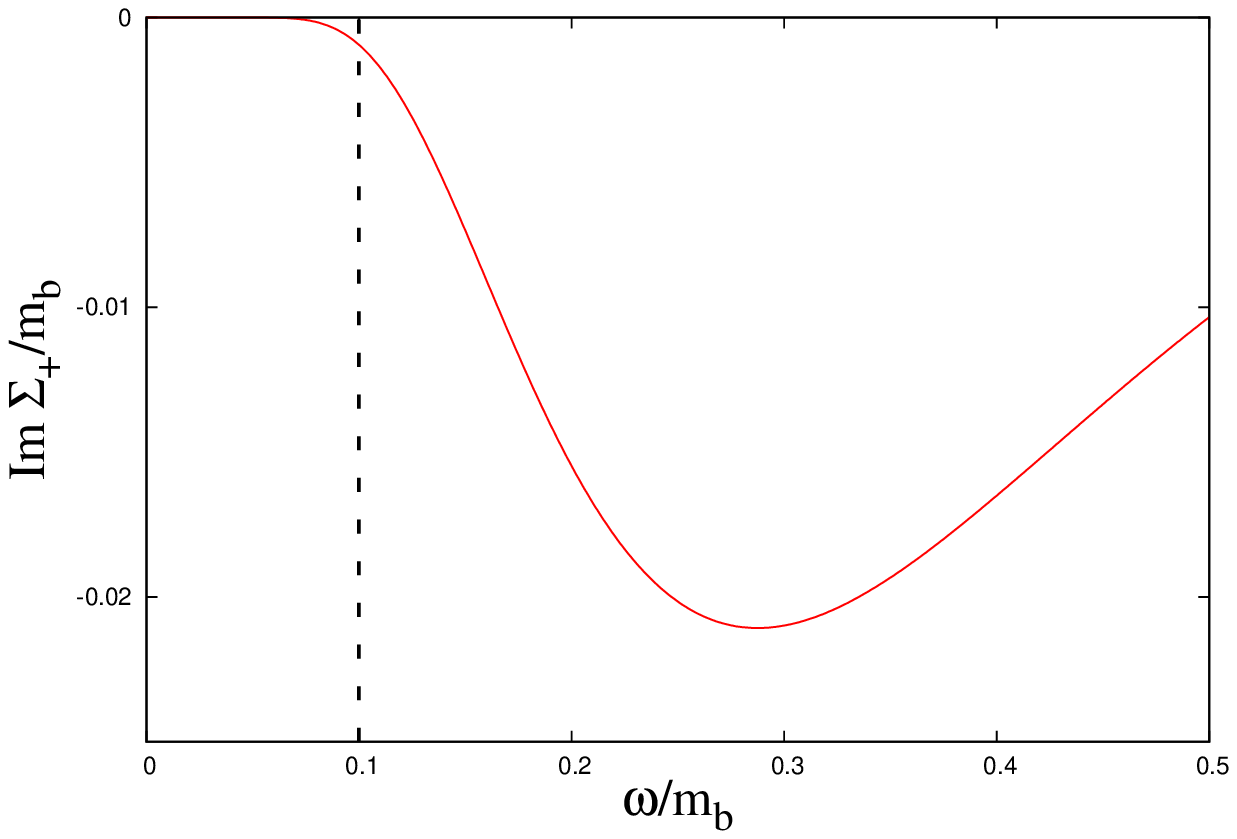}&
\includegraphics[clip,width=.32\textwidth]{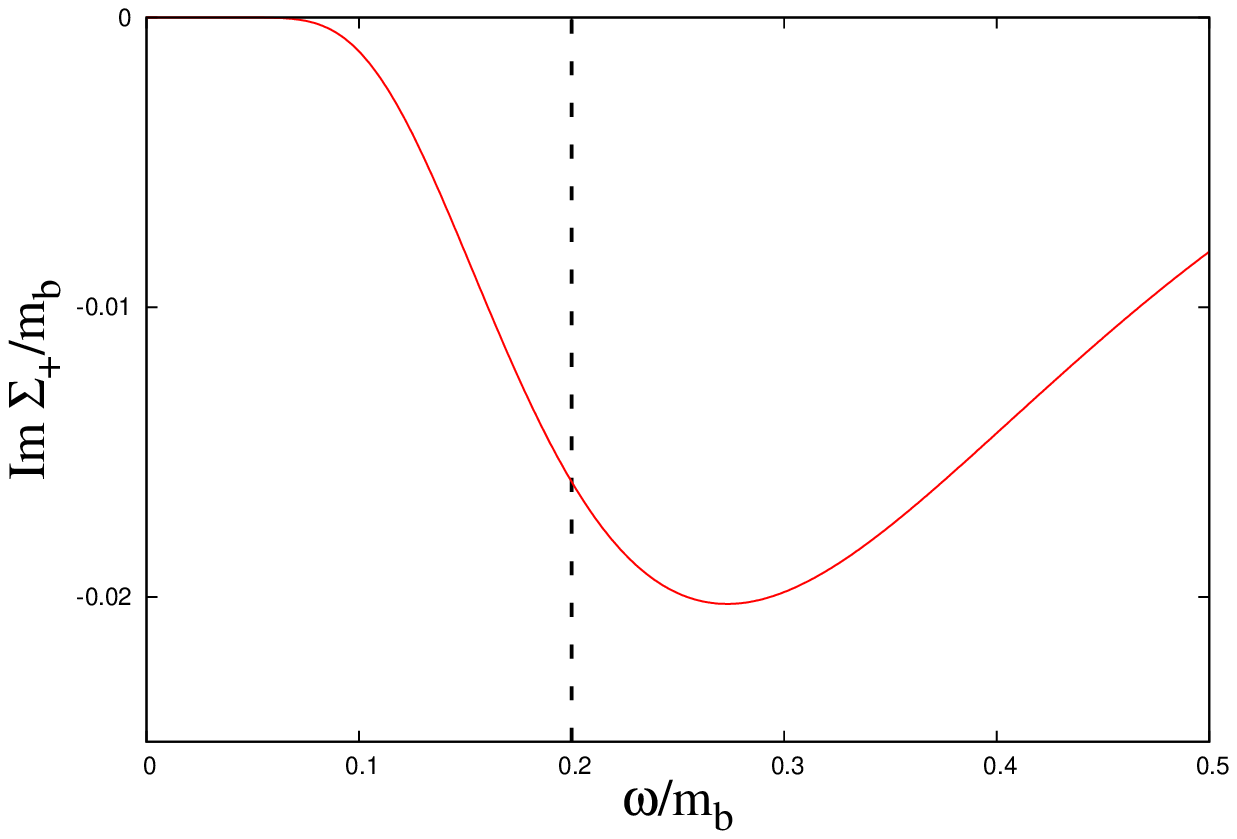}&
\includegraphics[clip,width=.32\textwidth]{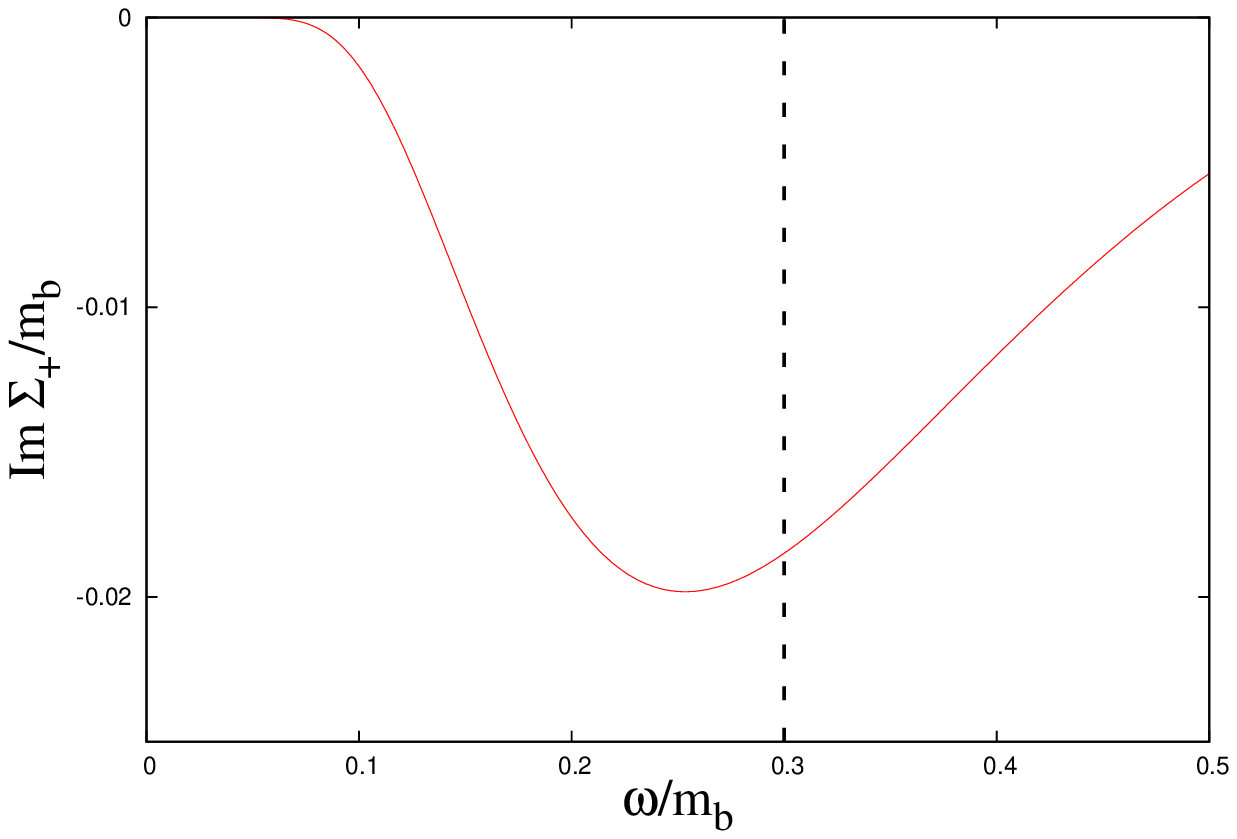}
\end{tabular}
\caption{The imaginary part of the 
quark self-energy ${\rm Im} \Sigma^{\rm S}_+(\omega)$ 
for $m_f/m_b=0.1,0.2$ and $0.3$ with fixed $\tau=0.5$.
The dotted lines represent $\omega=\mf$.
$\tau$ is chosen as the value at which $\rho_+(\omega)$
and the position of pole (A) start to show a clear deviation 
from those for $T=0$.}
\label{fig:se-t05}
\end{figure*}

Now, let us consider the level mixing for the present case
with $m_f\ne0$.
We first focus on the term (II) with the energy conservation
\begin{equation}
\omega_Q=-E_f(-{\bf k})+E_b({\bf k})
\equiv \omega^\mathrm{(II)}(k),
\label{eq:omega2}
\end{equation}
with $k=\vert {\bf k}\vert$,
where $\omega_Q$ is the energy of the quasi-quark $Q$ with zero 
momentum.
$\omega^\mathrm{(II)}(k)$ denotes the energy
of the two-particle state composed of a boson with momentum ${\bf k}$ and 
an antiquark-hole with momentum $-{\bf k}$, and is 
varied with $k$ from $M_{-}=m_b-m_f$ to 0 as shown by the
solid lines in Fig.~\ref{fig:level-mixing}.
Although $\omega^\mathrm{(II)}(k)$ can take any value
between $0$ and $M_{-}$, there exists a value
where the process corresponding to (II) is most probable.
Such a value may be given by the peak position of
Im$\Sigma_+^{\rm S}(\omega^{\rm(II)}(k))$ due to the term (II).
We denote the peak position as $\varepsilon_{\rm eff}^{\rm (II)}$ 
and show it in Fig.~\ref{fig:level-mixing}.
Then,
the physical energy spectrum at zero momentum is realized effectively
through a level repulsion between 
 $\varepsilon_{\rm eff}^{\rm (II)}$ 
and the on-shell free-quark energy $m_f$.
From Fig.~\ref{fig:level-mixing}, one finds that
$\varepsilon_{\rm eff}^{\rm (II)}$ gets to exceed $m_f$
at a mass between $m_f/m_b
=0.2$ and 0.3 when $m_f$ is increased.
This crossover is also seen from the movement of the peak position of 
Im$\Sigma_+^{\rm S}(\omega)$ relative to $m_f$, as shown
in Fig.~\ref{fig:se-t05}:
For $m_f/m_b=0.1$ and $0.2$, $\varepsilon_{\rm eff}^{\rm (II)}$ 
is larger than $m_f$, while for $m_f/m_b=0.3$, 
$\varepsilon_{\rm eff}^{\rm (II)}$ becomes
smaller than $m_f$.
Now, when $\varepsilon_{\rm eff}^{\rm (II)}>m_f$,
which is the case for $m_f/\mb < 0.21$,
the state with the unperturbed energy $m_f$ tends to have 
a lower energy than $m_f$,
while that with $\varepsilon_{\rm eff}^{\rm (II)}$ to
have higher energy than $\varepsilon_{\rm eff}^{\rm (II)}$
 by the coupling through the resonant scattering with the boson.
On the other hand, when $\varepsilon_{\rm eff}^{\rm (II)}<m_f$,
as is the case for $m_f/\mb>0.21$,
the state with the unperturbed energy $m_f$ tends to have 
a higher energy than $m_f$, while 
 that with $\varepsilon_{\rm eff}^{\rm (II)}$ to
have lower than $\varepsilon_{\rm eff}^{\rm (II)}$.
Thus we have seen that the level-crossing phenomenon between
poles (A) and (B) or the trajectory rearrangement 
at $m_f/m_b\simeq0.21$ can be nicely account for 
in terms of the level repulsion induced by the resonant scattering.

Similar discussions hold for the process of the term (III),
where the energy conservation reads
\begin{equation}
\omega_Q=E_f(-{\bf k})-E_b({\bf k})\equiv \omega^\mathrm{(III)}(k).
\label{eq:omega3}
\end{equation}
$\omega^\mathrm{(III)}(k)$ is drawn 
in Fig.~\ref{fig:level-mixing} with a solid line
in the negative energy region.
The most probable energy, $\varepsilon_{\rm eff}^{\rm (III)}$, in this case
is then negative, and thus $\varepsilon_{\rm eff}^{\rm (III)} < m_f$ is
satisfied irrespective of the value of $m_f$.
The physical energy spectrum at zero
momentum is again obtained effectively
as a result of the level repulsion between 
$\varepsilon_{\rm eff}^{\rm (III)}$ and $m_f$.
Because $\varepsilon_{\rm eff}^{\rm (III)}$ and $m_f$ never cross,
there is no qualitative change in the $T$-dependence of the
pole (C) unlike the process of the term (II):
The real part of the position of the pole (C)
monotonically decreases as $\tau$ increases,
as shown in Figs.~\ref{fig:pole01}, \ref{fig:pole02} and \ref{fig:pole03}.
As $m_f/m_b$ increases with $\tau$ fixed, 
$\varepsilon_{\rm eff}^{\rm (III)}$ goes farther from $m_f$,
which means that the effect of the level repulsion gets weaker
and thus the strength of the pole (C) gets weaker,
as shown in Fig.~\ref{fig:spc}.

We thus find that the $T$-dependence of the poles in the
quark propagator for each $m_f$ can be understood 
in terms of the level repulsion which is induced by 
a resonant scattering of the massive boson with a quark and
an anti-quark hole.

\section{Summary and Discussion}
\label{sec:summ}

In this paper, 
we have investigated the spectral properties of quarks
with a Dirac mass $m_f$ coupled with a scalar and pseudoscalar boson 
with a mass $m_b$ at finite temperature ($T$).
We have employed Yukawa models
to describe the system
and studied the spectral function and the pole structure of the quark
propagator for zero momentum at one loop.
The spectral function in the same models at finite $T$
was studied for the case with $m_f=0$ in Ref.~\cite{KKN2},
where the formation of the three-peak structure 
was observed at intermediate temperature $T/m_b\simeq1$.
In the present study, we have examined effects of $m_f$
on the spectral function in the range $m_f/m_b<0.3$ focusing on 
the spectral properties at zero momentum.
It was found that, although the three-peak structure
found in Ref.~\cite{KKN2} survives with finite $m_f$,
it tends to be suppressed as $m_f/m_b$ becomes larger.
We showed that the $T$ dependence of the free pole at $\omega=m_f$
changes qualitatively as $m_f/m_b$ becomes larger.

We have newly investigated the pole structure of the retarded
quark propagator in the lower-half complex-energy plane 
and the residues.
Our numerical calculation has identified the three poles  
corresponding to the peaks in the spectral function,
although near the imaginary axis,
there are infinite number of poles which do not form
peaks in the spectral function.
The $T$-dependence of the three poles is
consistent with that of the corresponding
peaks in the spectral function.
We have found that
the $T$-dependence of the poles changes qualitatively
around $m_f/m_b\simeq0.21$ ($0.23$) for Yukawa models
with scalar (pseudoscalar) boson.
This behavior is consistent with 
the behavior of the spectral function and
suggests that the physical contents of the collective
quark excitations at finite $T$ change qualitatively
around this value of the quark mass.
We have given an interpretation for this characteristic change
of the collective excitations in terms of the level crossing
induced by the resonant scattering \cite{KKN1,KKN2}.

The spectral function in the Yukawa model at finite $T$
was studied in Ref.~\cite{BBS92}, for the case with
$m_b=0$. In this case, the spectral function gradually
changes between a single- and two-peak structures
as $T/m_b$ is varied as an external parameter.
It is an interesting project to explore the pole structure
in this case, and the possible qualitative change of them
in the whole region of parameters $m_f$, $m_b$ and $T$.
This result will be reported elsewhere.

The present study  has a relevance to the study 
of effects of hadronic excitations \cite{HK85,Lattice}
in the QGP phase on the quasi-particle picture of the quark.
One of the plausible candidates of such a hadronic mode is 
the soft modes of the chiral transition \cite{HK85} 
since they can become light near $T_c$. 
The effect of the soft modes on the quark spectrum has been 
considered in Ref.~\cite{KKN1}
in the chiral effective model in the chiral limit.
In this case, the quarks are massless 
above $T_c$ and the soft modes become massless at $T_c$.
If the chiral symmetry is explicitly broken, on the other hand,
the chiral phase transition at finite $T$ becomes crossover
and the quarks have a finite mass for any value of $T$ \cite{NJL}.
The soft modes do not become massless, either.
These effects would suppress the formation of the three-peak
structure in the quark spectrum near $T_c$, 
as we have seen in the analysis of the present work.
The masses of quarks and soft modes, however, varies dynamically
as functions of $T$ in the QGP phase.
In order to investigate the effect of the soft modes 
more quantitatively,
the study of the quark spectrum with a chiral model incorporating
the explicit chiral symmetry breaking is desirable.

The quark spectral function near but above $T_c$ is recently 
analyzed in quenched lattice QCD in Ref.~\cite{Karsch:2007wc},
where the quark spectral function $\rho_+(\omega)$ 
at zero momentum is analyzed as a function of $m_f$.
It is shown that $\rho_+(\omega)$ is well described by 
a two-pole ansatz.
This result means that the three-peak structure
is hardly formed in the quark spectrum in quenched QCD.
We note, however, that the effect of the hadronic excitations 
is not incorporated in the quenched approximation.
The full lattice simulation 
would properly describe the
effect of such excitations on the quark spectrum.

It is an interesting problem to explore the effect of 
the multi-peak structure on experimental observables.
For example, the dilepton production rate is an interesting
candidate which can be affected by the structure of the
quark propagator \cite{BPY90}.
The excitation spectrum at low energy may modify the
thermodynamic quantities, such as the transport coefficients.
To examine these observables with the quark propagator
obtained in the present study is beyond the scope 
of this work and left as a future project.

Since the model which we have employed in this work is 
a simple but generic Yukawa model with a massive boson,
the results can be applied
to various systems at finite $T$
composed of a massive fermion and a massive boson.
For example, the quark spectrum might
have a multi-peak structure 
through couplings with massive bosonic states instead of
the soft modes, such as glueballs, charmonia, and so on.
Another example is the excitation spectra of neutrinos
coupled with weak bosons at $T$ near the masses of the weak 
bosons~\cite{Bo05}.
The multi-peak spectral function of
a Dirac particle with a vanishing or small mass might
be realized in some two-dimensional materials
such as graphene \cite{graphene} and an organic conductor 
$\alpha$-(BEDT-TTF)$_2$I$_3$ salt \cite{organic}
which exhibit Dirac-type dispersion relations.

  We are grateful to Berndt Muller, Kenji Fukushima
and the members of Hadron-Quark Seminar in Kyoto University
for their interest and encouragement.
  M. K. is supported by a Grant-in-Aid
  for Scientific Research by Monbu-Kagakusyo of Japan
  (No.~19840037).
  T. K. is supported by a Grant-in-Aid
  for Scientific Research by Monbu-Kagakusyo of Japan
  (No.~17540250).
  Y. N. is supported by a JSPS Grant-in-Aid for Scientific
  Research (\#18740140).
  This work is supported by the Grant-in-Aid for the 21st Century COE 
  ``Center for Diversity and Universality in Physics" of Kyoto
  University.
  The numerical calculations were carried out on 
  Altix3700 BX2 at YITP in Kyoto University.


\appendix
\section{Renormalization of the $T$-independent part
of the quark propagator}
\label{app:renormalization}

In this appendix, we calculate 
the $T$-independent part of the quark self-energy 
$\Sigma^R_{T=0}(\omega)$, which has an ultraviolet divergence.
The renormalization is carried out using the subtracted
dispersion relation.
In the following, we concentrate on the quark part 
$\Sigma_{+,T=0}(\omega)$, because the anti-quark part
$\Sigma_{-,T=0}(\omega)$
can be dealt with in the same way.

We first note that ${\rm Im} \Sigma_{+,T=0}$
is free from divergence
and thus can be calculated without regularization.
Taking a limit $T\rightarrow0^+$ in Eq.~\eqref{eq:imsigkin}, 
one obtains
\begin{align} 
\lefteqn{ {\rm Im} \Sigma_{+,T=0}(\omega) } \nonumber \\
&=-\frac{g^2}{32\pi}\frac{(\omega-\dm)(\omega+\tm)}{\omega^3}S(\omega)
\epsilon_\omega\theta(\omega^2-M_+^2),
\end{align}
where $\dm=\mb-\mf$, $\tm=\mb+\mf$, 
$S(\omega)=\sqrt{(\omega^2-\dm^2)(\omega^2-\tm^2)}$,
$\epsilon_\omega=\omega/\abs{\omega}$ is the sign of $\omega$ 
and $\theta(\omega)$ is the step function.

To determine the real part ${\rm Re} \Sigma_{+,T=0}(\omega)$,
we use the subtracted dispersion relation
\begin{equation}
\begin{split}
\re \Sigma_{+,T=0}(\omega)
=&\sum_{l=0}^{n-1}\frac{(\omega-\alpha)^l}{l!}c_l\\
&+\frac{(\omega-\alpha)^n}{\pi}\mathcal{P}
\int_{-\infty}^{\infty}d\zeta
\frac{\imsigpvac(\zeta)}{(\zeta-\omega)(\zeta-\alpha)^n},
\end{split}
\label{eq:nKK}
\end{equation}
where $\alpha$ denotes the renormalization point,
$c_l$ are the subtraction constants,
and $n$ is the number of subtraction.
To regularize the ultraviolet divergence, we use Eq.~(\ref{eq:nKK})
with $n=2$.

The subtraction constants $c_0$ and $c_1$ may be  determined 
by the on-shell renormalization conditions for
the mass  and the wave function  at $\alpha=\mf$,
which are tantamount to requiring that
\begin{equation}
\Sigma_{+,T=0}(\mf) = 0,
\label{eq:renorm1}
\end{equation}
and 
\begin{equation}
\left.\frac{\partial \Sigma_{+,T=0}(\omega)}{\partial \omega}
\right|_{\omega=\mf}=0,
\label{eq:renorm2}
\end{equation}
respectively.
From Eqs.~(\ref{eq:renorm1}) and (\ref{eq:renorm2}),
we obtain $c_0=c_1=0$, and then the renormalization is completed.

\section{Generic nature of 
 the three-peak structure in the quark spectrum; an analytic toy model}
\label{app:3peak}

\begin{figure}[tn]
\begin{center}
\includegraphics[clip,width=.48\textwidth]{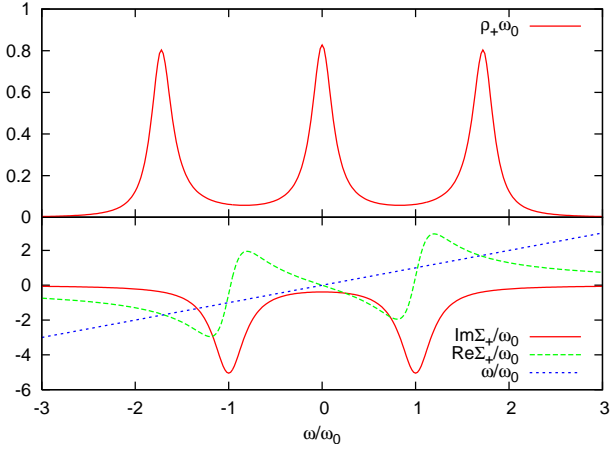}
\end{center}
\caption{The spectral function $\rho_+$ and the self-energy $\Sigma_+$
for $\gamma/\omega_0=0.2$. The quark mass is taken to be zero.}
\label{fig:lor02}
\end{figure}

We have seen in the main text and Refs.~\cite{KKN1, KKN2}
that the two-peak structure in the imaginary part of the
quark self-energy leads
to  the three-peak structure in the 
quark spectral function.
In this appendix, we show that this has a generic nature
using a simple but generic model.
For simplicity, we consider the case for vanishing momentum as in the main
text. 

First, let us consider the quark spectrum for the massless quark.
To investigate effects of the peak structure in the imaginary part
of the self-energy on the quark spectrum,
we assume that the imaginary part of the retarded quark self-energy consists
of two peaks expressed with Lorentzian forms,
\begin{equation}
  \textrm{Im}\Sigma_+(\omega) = 
  -M^2 \bigg[ \frac{\gamma_1}{(\omega_1-\omega)^2+\gamma_1^2}
  +\frac{\gamma_2}{(\omega_2-\omega)^2+\gamma_2^2} \bigg],
  \label{im2}
\end{equation}
where $\omega_{1,2}$ denote the peak positions and $\gamma_{1,2}$ denote
the widths.
An overall factor $M^2$ having mass dimension two is put
because the quark self-energy has mass dimension one.
Since we are interested in the number of peaks in the quark spectrum,
the deviation of the peak shape from the Lorentzian form
is not important here.
For the massless quark, we have $\omega_0\equiv\omega_1=-\omega_2$ and
$\gamma\equiv\gamma_1=\gamma_2$ from a symmetry property of the self-energy.
As discussed in the main text and Refs.~\cite{KKN1, KKN2},
the formation of the peaks in Im$\Sigma_+$ is caused by the Landau damping
which is scattering processes with a boson.
As $T$ increases, the peaks become higher rapidly as shown 
in Figs.~\ref{fig:se00} and \ref{fig:se01}, which corresponds to
$\gamma/\omega_0$ being smaller.

The corresponding real part of the self-energy is uniquely determined
from the imaginary part:
Because the Landau damping is a medium effect, the real part
is ultraviolet finite and given by the un-subtracted dispersion relation,
\begin{align}
  \textrm{Re}\Sigma_+(\omega) &= \mathcal{P}\int_{-\infty}^\infty
  \frac{d\omega'}{\pi} \frac{\textrm{Im}\Sigma_+(\omega')}{\omega'-\omega}
  \notag \\
  &= M^2\bigg[\frac{\omega-\omega_0}{(\omega_0-\omega)^2+\gamma^2}+
     \frac{\omega+\omega_0}{(\omega_0+\omega)^2+\gamma^2}\bigg].
\end{align}
The spectral function is then obtained from the self-energy,
$\rho_+(\omega)= -1/\pi \cdot \textrm{Im}\Sigma_+(\omega)/
[(\omega-\textrm{Re}\Sigma_+(\omega))^2+\textrm{Im}\Sigma_+(\omega)^2]$.

As an example, we plot the spectral function and the self-energy
for $\gamma/\omega_0=0.2$ in Fig.~\ref{fig:lor02}.
For simplicty, we have set $M=\omega_0$.
From the number of crossing points between 
Re$\Sigma_+$ and the line $\omega/\omega_0$, one finds that
this self-energy is close to that for $T=1.5$ shown in Fig.~\ref{fig:se00}.
We see that the resultant quark specral function $\rho_+$ has a clear
three-peak structure.
For smaller values of $\gamma/\omega_0$, i.e. higher $T$,
the three peaks are sharper and higher than those in Fig.~\ref{fig:lor02}.

For very high $T$, however, the strength of the central peak 
in $\rho_+$ gets weaker and eventually $\rho_+$ becomes to
have only two peaks which correspond to the normal 
quasi-quark and the anti-plasimino as shown in Fig.~\ref{fig:spc}.
In the Yukawa models,
this can be understood from the fact that as $T$ increases,
in addition to two peaks in Im$\Sigma_+$ being higher,
the positions of the two peaks approaches the origin.
In terms of Eq.~(\ref{im2}), this means 
$\omega_1/T,\omega_2/T\to 0$ for $T\to\infty$, thus
it becomes hard to distinguish the two peaks, and thus Im$\Sigma_+$ 
only shows 
a one-peak structure as a whole.
One can show in a generic way that
the one-peak structure of Im$\Sigma_+$ leads to
a two-peak structure of the quark spectral function.
In fact,
this rule can be extended to the cases where
Im$\Sigma_+$ has multiple peaks;
the number of the peaks in Im$\Sigma_+$ 
is significantly reflected
in the number of the peaks in $\rho_+$.

\begin{figure}[ht]
\begin{center}
\includegraphics[clip,width=.48\textwidth]{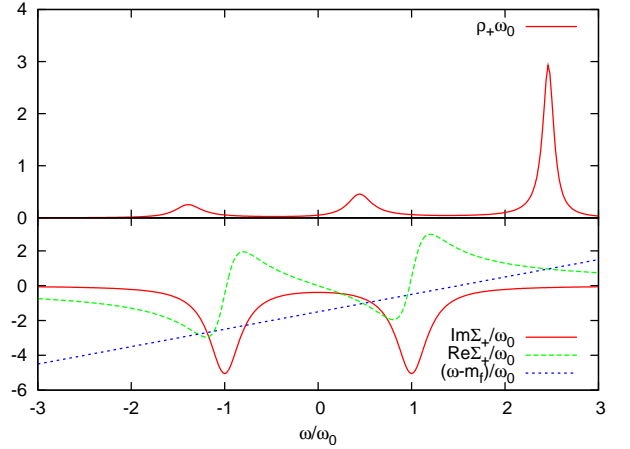}
\end{center}
\caption{The spectral function $\rho_+$ and the self-energy $\Sigma_+$
for $\gamma/\omega_0=0.2$ and $m_f/\omega_0=1.5$.}
\label{fig:lor02m}
\end{figure}

Next we consider the quark spectrum for the massive quark.
In this case, $\omega_1$ and $\gamma_1$ are in general not identical to
$-\omega_2$ and $\gamma_2$, respectively.
However, from Fig.~\ref{fig:se01},
 we see that such asymmetry is not so large for small
values of the quark mass $m_f$.
Then, for simplicty, we employ the same self-energy as that for the massless 
quark.
The mass effect only enters the expression of the spectral function,
$\rho_+(\omega)= -1/\pi \cdot \textrm{Im}\Sigma_+(\omega)/
[(\omega-m_f-\textrm{Re}\Sigma_+(\omega))^2+\textrm{Im}\Sigma_+(\omega)^2]$.

As an example, we plot the spectral function and the self-energy
for $\gamma/\omega_0=0.2$ and $m_f/\omega_0=1.5$ in Fig.~\ref{fig:lor02m}.
From the number of crossing points between 
Re$\Sigma_+$ and the line $(\omega-m_f)/\omega_0$,
this self-energy is colse to that for $T=1.5$ in Fig.~\ref{fig:se01}.
We see that the rightmost peak is enhanced, which results from
the shift of the crossing points.
This result of the quark spectrum is also consistent with that in 
Fig.~\ref{fig:spc}.

For very high $T$, the effect of the quark mass becomes relatively
 smaller and thus the quark spectrum 
approaches that for the massless quark mentioned above.

In short, we have shown that 
a simple but plausible assumption on the peak structure in the imaginary
part of the self-energy generically leads to  
the multi-peak structure in the quark spectral function
obtained in dynamical models such as the  Yukawa models.


\end{document}